\documentclass[submission, Phys]{SciPost}
\usepackage{grffile}
\usepackage{lineno}
\usepackage{graphicx}  % needed for figures
\usepackage{dcolumn}   % needed for some tables
\usepackage{bm}        % for math
\usepackage{amssymb}   % for math
\usepackage{amsmath}
\usepackage{blkarray, multirow, graphicx, diagbox, color, xcolor, colortbl}
\usepackage{bbm, bbold}
\usepackage{glossaries}
\usepackage{hyphenat}
\usepackage{ifthen}
\usepackage{xkeyval}
\usepackage{moreverb}
\usepackage{rotating}
\usepackage{wrapfig}
\usepackage{slashbox}
\usepackage{xspace}
\usepackage{nicefrac}
\usepackage[]{units}
\usepackage{physics}
\usepackage{braket}
\usepackage[inline]{enumitem}
\usepackage{tabto}
\usepackage{listings}
\usepackage{xstring}
\def\ReplaceStr#1{%
	\IfSubStr{#1}{p}{%
		\StrSubstitute{#1}{p}{.}}{#1}}

\usepackage[style=base]{caption}
\usepackage{subcaption}
\usepackage{booktabs}
\newcommand\subfigref[1]{(\protect\subref{#1})}

\usepackage{tikz}
\usepackage{calc}
\usetikzlibrary{external}
\usepackage{pgffor}
\usepackage{pgfplots}
\pgfplotsset{compat=1.13}
\usepackage{pgfplotstable}
\usepgfplotslibrary{groupplots}
\usepgfplotslibrary{fillbetween}

\tikzstyle{n} = [draw,shape=ellipse,minimum size=1.5em,inner sep=0pt,fill=white!20, minimum width=2.5em]
\tikzstyle{Init} = [n,color=green,fill=green!20,text=black]
\tikzstyle{Fin} = [n,color=red,fill=red!20,text=black]
\tikzstyle{Ghost} = [minimum size=1.5em,inner sep=0pt,color=white,text=black]
\tikzstyle{Multiple} = [draw,shape=rect,minimum size=2em,inner sep=0pt]

\tikzstyle{ghostA} = [text=red!70,thick, minimum size=2*(5pt-\pgflinewidth), inner sep=0pt, outer sep=0pt]
\tikzstyle{ghostB} = [text=blue!70,thick, minimum size=2*(5pt-\pgflinewidth), inner sep=0pt, outer sep=0pt]
\tikzstyle{siteA} = [regular polygon, regular polygon sides=3, shape border rotate= 30, draw=red!50,fill=red!20,thick,inner sep=0pt,minimum width=1.5em,font=\footnotesize]
\tikzstyle{siteB} = [regular polygon, regular polygon sides=3, shape border rotate= -30, draw=green!50,fill=green!20,thick,inner sep=0pt,minimum width=1.5em,font=\footnotesize]
\tikzstyle{op} = [regular polygon, regular polygon sides=4, draw=orange!50, fill=orange!20, thick, inner sep=0.2pt, minimum width=1.25em, minimum height=1.5em,font=\footnotesize]
\tikzstyle{opghost} = [regular polygon, regular polygon sides=4, thick, inner sep=0.2pt, minimum width=1.25em, minimum height=1.5em,font=\footnotesize]
\tikzstyle{site} = [circle,draw=blue!50,fill=blue!20,thick,inner sep=0.2pt,minimum width=1.25em,font=\footnotesize]
\tikzstyle{hiddensite} = [circle,draw=white!50,fill=white!20,thick,inner sep=0.2pt,minimum width=1.25em,font=\footnotesize]
\tikzstyle{nosite} = [circle,draw=white,fill=white,thick,inner sep=0.1pt,minimum width=1.5em]
\tikzstyle{ghost} = [font=\footnotesize]
\tikzstyle{intersite} = [regular polygon, regular polygon sides=4, shape border rotate= 45, draw=black!50,fill=black!20,thick,inner sep=0pt,minimum width=1.5em]
\tikzstyle{ld} = [inner sep=1pt, font=\small]
\tikzstyle{unsite} = [circle, outer sep=0pt,inner sep=0.2pt,minimum width=1.25em]

\definecolor{colorA}{rgb} {0.58,0,0.8275}
\definecolor{colorB}{rgb} {0.11,0.663,0.51}
\definecolor{colorC}{rgb} {0.3373,0.7059,0.9137}
\definecolor{colorD}{rgb} {0.902,0.6235,0}
\definecolor{colorE}{rgb} {0.9451,0.902,0.3255}
\definecolor{colorF}{rgb} {0.3373,0.3255,0.902}
\definecolor{colorG}{rgb} {0.9451,0.3255,0.3373}

\usetikzlibrary
{
	calc,
	decorations,
	plotmarks,
	patterns,
	positioning,
	petri,
	arrows,
	decorations.markings,
	backgrounds,
	fit,
	graphs,
	shapes.geometric,
	decorations.pathmorphing,
	shapes.misc,
	shapes,
	tikzmark,
	pgfplots.colorbrewer,
	fpu,
}

\usepackage{ocgx}

\usetikzlibrary{ocgx}

\pgfplotsset{
        cycle from colormap manual style/.style={
            x=3cm,y=10pt,ytick=\empty,
            stack plots=y,
            every axis plot/.style={line width=2pt},
        },
}

\tikzset{->-/.style={decoration={
			markings,
			mark=at position .5 with {\arrow{>}}},postaction={decorate}}}

\tikzset{-<-/.style={decoration={
			markings,
			mark=at position .5 with {\arrow{<}}},postaction={decorate}}}

\tikzstyle{orientedsnake} = [
decorate, 
decoration={snake},
->
]  
\tikzstyle{orientedshortarrow} = [
decoration={markings,
	mark=at position .33 with {\arrow{>}}},
postaction={decorate}
]  
\tikzstyle{orientedlongarrow} = [
decoration={markings,
	mark=at position .67 with {\arrow{>}}},
postaction={decorate}
]
\tikzset{dbl/.style={double,
		double equal sign distance,
		-implies,
		shorten >=10pt,
		shorten <=10pt}}
\tikzset{
	between/.style args={#1 and #2}{
		at = ($(#1)!0.5!(#2)$)
	}
}

\newcommand{\nodagger}[0]{{\phantom{\dagger}}}

\pgfmathdeclarefunction{peierlspotential}{6}%
{% 	plot with: \addplot[black] {peierls_potential(R, e0, k0, t0, c, s)};
	\pgfmathparse
	{
		#2 / (#3 * #5) 
		* sin(deg(#1 * #3 - #3 * #5 * (x)))
		* exp(-( ( #1 - #5 * ((x) - #4) ) )^2.0 / ( #6^2.0 ) )
	}%
}

\pgfmathdeclarefunction{linearFct}{2}%
{%
	\pgfmathparse{#1*x+#2}%
}
\pgfmathdeclarefunction{quadFct}{3}%
{%
	\pgfmathparse{#1*x*x+#2*x+#3}%
}
\pgfmathdeclarefunction{logFct}{3}%
{%
	\pgfmathparse{#1*ln(#3*x)+#2}%
}
\pgfmathdeclarefunction{expFct}{2}%
{%
	\pgfmathparse{#1*exp(-x/#2)}%
}

\newboolean{buildCurrentBlockInline}
\setboolean{buildCurrentBlockInline}{false}
\newboolean{rebuildCurrentBlock}
\setboolean{rebuildCurrentBlock}{false}

\newboolean{rebuildData}
\setboolean{rebuildData}{false}

\newboolean{removeData}
\setboolean{removeData}{false}

\graphicspath{{figures/autogen/}}
	
\newboolean{buildtikzpics}
\setboolean{buildtikzpics}{false}
\newif\ifrebuildtikz
\newif\ifChangeMode
\ChangeModetrue
\ChangeModefalse
\ifthenelse{\boolean{buildtikzpics}}
{
	\rebuildtikztrue
	\usetikzlibrary{external}
	\tikzexternalize[optimize=false,prefix=figures/autogen/]
}
{
	\rebuildtikzfalse
}

\usepackage{ulem}
\usepackage{cleveref}
\Crefname{appendix}{Appendix}{Appendices}
\Crefname{equation}{Equation}{Equations}
\Crefname{figure}{Figure}{Figures}
\Crefname{section}{Section}{Sections}
\Crefname{tabular}{Tabular}{Tabulars}
\crefname{appendix}{App.}{Apps.}
\crefname{equation}{Eq.}{Eqs.}
\crefname{figure}{Fig.}{Figs.}
\crefname{section}{Sec.}{Secs.}
\crefname{tabular}{Tab.}{Tabs.}

\ExplSyntaxOn
\DeclareExpandableDocumentCommand \eval { m } { \fp_eval:n { #1 } }
\ExplSyntaxOff

\makeatletter
\def\pgfplotsutil@decstringcounter#1{%
 \begingroup
  \c@pgf@counta=#1\relax
  \advance\c@pgf@counta by -1
  \edef#1{\the\c@pgf@counta}%
  \pgfmath@smuggleone#1%
 \endgroup
}%

\pgfplotsset{
/pgfplots/each nth point*/.style 2 args={%
/pgfplots/x filter/.append code={%
 \ifnum\coordindex=0
  \def\c@pgfplots@eachnthpoint@xfilter{0}%
  \edef\c@pgfplots@eachnthpoint@xfilter@cmp{#1}%
 \else
  \ifnum\coordindex>#2\relax
   \pgfplotsutil@advancestringcounter\c@pgfplots@eachnthpoint@xfilter
   \ifx\c@pgfplots@eachnthpoint@xfilter@cmp\c@pgfplots@eachnthpoint@xfilter
    \def\c@pgfplots@eachnthpoint@xfilter{0}%
   \else
    \let\pgfmathresult\pgfutil@empty
   \fi
  \fi
 \fi
}%
},
}
\makeatother
\newacronym{PCMO}{PCMO}{praseodymium\hyp calcium\hyp manganite}
\newacronym{BCS}{BCS}{Bardeen\hyp Cooper\hyp Schrieffer}
\newacronym{1D}{1D}{one\hyp dimensional}
\newacronym{2D}{2D}{two\hyp dimensional}
\newacronym{3D}{3D}{three\hyp dimensional}
\newacronym{MPS}{MPS}{matrix\hyp product state}
\newacronym{MPO}{MPO}{matrix\hyp product operator}
\newacronym{SVD}{SVD}{singular\hyp value decomposition}
\newacronym{QCS}{QCS}{quantum\hyp computer simulator}
\newacronym{QC}{QC}{quantum computer}
\newacronym{FSM}{FSM}{finite\hyp state machine}
\newacronym{ACA}{ACA}{adaptive cross\hyp approximation}
\newacronym{CDW}{CDW}{charge\hyp density wave}
\newacronym{SDW}{SDW}{spin\hyp density wave}
\newacronym{ARPES}{ARPES}{angle-resolved photoemission spectroscopy}
\newacronym{OBC}{OBC}{open-boundary conditions}
\newacronym{PBC}{PBC}{periodic-boundary conditions}
\newacronym{TEBD}{TEBD}{time-evolving block decimation}
\newacronym{2TDVP}{2TDVP}{two-site time-dependent variational principle}
\newacronym{iff}{iff}{if and only if}
\newacronym{DFT}{DFT}{density\hyp functional theory}
\newacronym{DMFT}{DMFT}{dynamical mean\hyp field theory}
\newacronym{DMRG}{DMRG}{density\hyp matrix renormalization group}
\newacronym{QMC}{QMC}{quantum Monte Carlo}
\newacronym{AIM}{AIM}{Anderson impurity model}
\newacronym{SIAM}{SIAM}{single impurity Anderson model}
\newacronym{LDA}{LDA}{local\hyp density approximation}
\newacronym{LBNL}{LBNL}{Lawrence Berkeley National Laboratory}
\newacronym{VQE}{VQE}{variational\hyp quantum eigensolver}
\newacronym{ED}{ED}{exact diagonalization}
\newacronym{QPT}{QPT}{quantum phase transition}
\newacronym{QCP}{QCP}{quantum critical point}
\newacronym{ETH}{ETH}{eigenstate thermalization hypothesis}
\newacronym{AKLT}{AKLT}{Affleck\hyp Lieb\hyp Kennedy\hyp Tasaki}
\newglossaryentry{QR}{name={QR},description={QR decomposition}}
\newacronym{TNS}{TNS}{tensor\hyp network state}
\newacronym{PS}{PS}{pseudo site}
\newacronym{ppDMRG}{ppDMRG}{projected purified \gls{DMRG}}
\newacronym{LBO}{LBO}{local\hyp basis optimization}
\newacronym{SC}{SC}{superconducting}
\newacronym{1RDM}{1RDM}{single\hyp site reduced density\hyp matrix}
\newacronym{MF}{MF}{Mean\hyp Field}

\begin{document}
\begin{center}
	{\Large 
		\textbf{
		Transient superconductivity in three-dimensional Hubbard systems by combining matrix product states and self-consistent mean-field theory
		}
	}
\end{center}

\begin{center}
	S.~Marten\textsuperscript{1},
	G.~Bollmark\textsuperscript{2},
	T.~K\"ohler\textsuperscript{2},
	S.R.~Manmana\textsuperscript{1},
	A.~Kantian\textsuperscript{2,3}
\end{center}

\begin{center}
{\bf 1} 
Institut f\"ur Theoretische Physik, Georg\hyp August\hyp Universit\"at  G\"ottingen, 37077 G\"ottingen, Germany
{\bf 2} 
Department of Physics and Astronomy, Uppsala University, Box 516, S-751 20 Uppsala, Sweden
{\bf 3} 
SUPA, Institute of Photonics and Quantum Sciences, Heriot-Watt University, Edinburgh EH14 4AS, United Kingdom
\end{center}
\section*{Abstract}
{
	We combine \gls{MPS} and \gls{MF} methods to model the real\hyp time evolution of a \gls{3D} extended Hubbard system formed from \gls{1D} chains arrayed in parallel with weak coupling in-between them. 
	This approach allows us to treat much larger 3D systems of correlated fermions out\hyp of\hyp equilibrium over a much more extended real\hyp time domain than previous numerical approaches. 
	We deploy this technique to study the evolution of the system as its parameters are tuned from a \gls{CDW} phase into the \gls{SC} regime, which allows us to investigate the formation of transient non\hyp equilibrium \gls{SC}.
	In our ansatz, we use \gls{MPS} solutions for chains as input for a self\hyp consistent time\hyp dependent \gls{MF} scheme. 
	In this way, the \gls{3D} problem is mapped onto an effective \gls{1D} Hamiltonian that allows us to use the \gls{MPS} efficiently to perform the time evolution, and to measure the \acrshort{BCS} order parameter as a function of time.
	Our results confirm previous findings for purely \gls{1D} systems that for such a scenario superconductivity forms in a transient state. 
}

%
% Guideline: if your paper is longer than 6 pages, include a TOC
\vspace{10pt}
\noindent\rule{\textwidth}{1pt}
\tableofcontents\thispagestyle{fancy}
\noindent\rule{\textwidth}{1pt}
\vspace{10pt}

\section{\label{sec:introduction}Introduction}
\Gls{SC} inspires researchers since its discovery in 1911 by H. K. Onnes.
Its explanation is a true challenge, and it took more than 40 years to introduce the meaningful theoretical \gls{BCS} framework to explain the findings by a suitable \gls{MF} theory. 
In the 1980s, \gls{SC} at high critical temperatures $T_c$ \cite{bednorz1986_possibleHighTcSC, takagi1987hightTc_SC_cuprates_experiment, schilling1993highTc_SC_cuprates_experiment,  dagotto1994_correlatedElectronsInHighTcSC} was discovered, which seemed not to be described by \gls{BCS} theory.
In facts, its theoretical description presents an still ongoing challenge.
It is believed that strongly correlated electron motion is the underlying reason for this type of \gls{SC} state.
Many\hyp body models such as the Hubbard-\cite{hubbard1963electronCorrelation_in_narrow_Ebands,Gutzwiller1963_HubbardModel2,kanamori1963HubbardModel3,essler2005TheOneDimHubbardModel,fazekas,auerbach2012HubbardModelandHeisenbergModel} or the  $t$-$J$-model\cite{tJoriginal1,tJoriginal2,auerbach2012HubbardModelandHeisenbergModel,dagotto1994_correlatedElectronsInHighTcSC,tJ1977} have been investigated to study this question.
In more recent developments, experiments claimed to have detected metastable, light\hyp induced \gls{SC} states after pushing materials out\hyp of\hyp equilibrium in pump\hyp probe setups.
Such a transient non\hyp equilibrium \gls{SC} regime is possibly even detected above the equilibrium critical temperature $T_c$ \cite{fausti_lightInducedSC,buzzi2020_photomolecularSC, kaiser2014_opticallyInducedSC, mitrano2016possibleLightInducedSC}. 
On the theoretical side, this scenario has been studied in various approaches, e.g., numerically \cite{Echstein_2010_InteractionQuench_HubbardModel, Paeckel_superconductivity, bittner2019possibleLightInducedSC_HubbardModel_phaseDiagram}, but many basic question about the mechanisms remain open. 
While many experiments rely on the time\hyp dependent optical conductivity as a probe for nonequilibrium \gls{SC}, Paeckel et al.~\cite{Paeckel_superconductivity} recently showed that this measure lacks specificity for \gls{SC} order, at least in the numerically setup studied.
This setup consists of a quantum quench on a purely \gls{1D} extended Hubbard system using \gls{MPS}. 
That work suggests alternative measurements, which would be better suited to detecting the onset of the \gls{SC} state in the dynamically evolving system. 
However, while this \gls{MPS} approach is unbiased and highly accurate, it is so far largely restricted to \gls{1D} systems, especially when treating out\hyp of\hyp equilibrium dynamics. 
The question is thus if the findings of Paeckel et al.~are specific to \gls{1D}, with its strong quantum and thermal fluctuations, or whether their results also apply to the realm of higher dimensional systems. 
This sets an immediate challenge: which theoretical method could address the dynamics of interacting fermions out\hyp of\hyp equilibrium in \gls{3D}? 
On their own, even in \gls{1D}, \gls{MPS} methods may require exponentially increasing resources as simulation time grows in order to maintain a set accuracy. 
This is due to the strong growth in bipartite entanglement in these systems with time: for \gls{MPS} approaches to be efficient, this entanglement should not be too large. 
Furthermore, already for equilibrium calculations long\hyp range interactions, which are needed to represent \gls{2D} and \gls{3D} systems in \gls{1D}, increase the entanglement dramatically.
Hence, the time evolution of generic \gls{2D} and \gls{3D} systems are entirely out of reach for \gls{MPS}.
However, at large spatial dimensions, real\hyp time non\hyp equilibrium \gls{DMFT} is a powerful approach \cite{georges_RMP,Review_tDMFT}. 
In these approaches, one or a few lattice sites - the impurity or, respectively, the cluster - are retained explicitly, including all interactions of the original, infinitely\hyp large lattice. 
In \gls{DMFT}, the effect of this remainder\hyp lattice on the cluster is mimicked via a free\hyp electron bath that is coupling to it. 
The parameters of this bath are fixed via self\hyp consistency conditions. 
Solving these cluster\hyp bath systems within this self\hyp consistency constraint is typically achieved by applying \gls{QMC} techniques in the real time domain. 
These techniques suffer from a strong sign\hyp problem, i.e., their numerical error grows exponentially as the cluster\hyp size and the real\hyp time domain, over which the simulation runs, are increased. 
In practice, a few sites and time scales on the order of the electron tunneling are accessible. 
Alternatively, \gls{MPS} solvers can be used within such real\hyp time non\hyp equlibrium \gls{DMFT}; however, due to the long\hyp range tunneling in these systems between bath and cluster sites, and the strong growth of entanglement with time, these will also be limited to a few sites and short times. 
This leads us to the scope of the present paper: with current methods it seems practically impossible to perform meaningful simulations of dynamically\hyp induced \gls{SC} in a \gls{3D} system. 
For \gls{MPS} methods, the growth of entanglement with system size and simulation time is prohibitive, for non\hyp equilibrium real\hyp time \gls{DMFT}, the large clusters and long times required to resolve the onset of a potentially weak \gls{SC} order appear out of reach. 
However, as we demonstrate in the following, it is possible to make such cases treatable with \gls{MPS} techniques employing a static \gls{MF} ansatz provided that the spectrum has a large energy gap.
In this way, it is possible to capture strong correlations by the \gls{MPS}, and treat the full \gls{3D} system more accurately than by applying a pure \gls{MF} treatment. 
Indeed, related approaches have been studied before at equilibrium \cite{PhysRevB.83.054407}, where at least qualitative behavior was reproduced correctly compared to appropriate \gls{QMC} simulations~\cite{Gunnar_coupled_chains_electrons,Gunnar_coupled_chains}. 
In these approaches, weakly coupled chains or ladders are stacked up into \gls{3D} cubic systems, which thus have anisotropic tunneling --- much stronger inside the \gls{1D} systems than in\hyp between them in the two orthogonal directions. 
For the case of fermions, the \gls{MF} approximation can be introduced if each of the constituent \gls{1D} systems has a gapped energy sector, such as a spin gap, and thus single\hyp fermion tunneling in\hyp between \gls{1D} systems is suppressed in this weak\hyp coupling regime~\cite{Gunnar_coupled_chains_electrons}. 
Just as for the equilibrium case~\cite{Gunnar_coupled_chains_electrons}, it is this crucial ingredient that allows us to perform real\hyp time evolution for a much higher number of correlated sites than non\hyp equilibrium real\hyp time \gls{DMFT}, as well as extending the real\hyp time domain enough to perform a meaningful simulation of the dynamically-induced \gls{SC} in a \gls{3D} system. 
Within this well\hyp behaved domain, we apply our real\hyp time \gls{MPS}+\gls{MF} technique to study the time\hyp evolution of the \gls{BCS} order parameter after fast ramping the system from an insulating starting state into a parameter regime where the system would be \gls{SC} in equilibrium. 
As a consequence, we observe the onset of a non\hyp equilibrium \gls{SC} state.
The paper is structured as follows: In \cref{sec:model}, we recapitulate the \gls{MF} ansatz for weakly coupled Hubbard chains used in equilibrium, developed originally in \cite{Gunnar_coupled_chains_electrons}.
In \cref{sec:algorithm}, we introduce the extension to a self\hyp consistent time\hyp dependent \gls{MPS}+\gls{MF} scheme to study the time evolution of a \gls{3D} extended Hubbard system, which consists of weakly coupled chains.
In \cref{sec:resultsAndDiscussion}, we present our results for the \gls{BCS} order parameter and a detailed discussion of the convergence behavior of the method when treating \gls{3D} arrays formed from chains, each up to ${L=30}$ lattice sites long.
The time evolution of the \gls{SC} order parameter shows indeed that in both finite systems as well as the thermodynamic limit a transient \gls{SC} state can be entered.
We further analyze the dependence of our results on the parameters of the simulations. 
In \cref{sec:conclusion} we conclude and give an outlook to possible further developments and applications of our method.
The appendices discuss further details on the method at equilibrium, as well as further details of the simulations out\hyp of\hyp equilibrium. 
\section{\label{sec:model}Mapping of the \gls{3D} system onto a \gls{1D} self\hyp consistent chain}
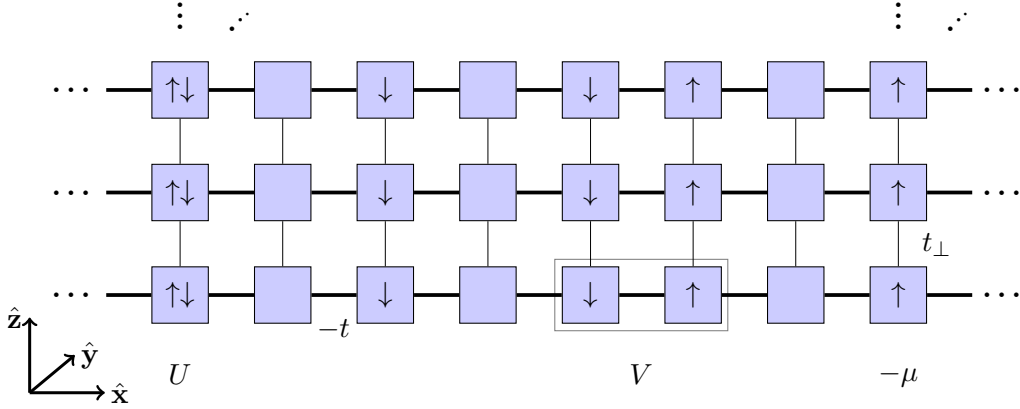
\begin{figure}[!t]
	\centering
	\ifthenelse{\boolean{buildtikzpics}}%
	{%
		\tikzsetnextfilename{model}
\begin{tikzpicture}[
site/.style={rectangle, draw=black, fill=blue!20, align=center, minimum size = 0.75cm},
ghost/.style={align=center},
box/.style={rectangle, draw=gray, minimum height = 0.95cm, minimum width = 2.3cm}
]
    \begin{scope}[node distance=0.6cm]
		
		\node (x) at (-0.8,-1.3) {$\hat{\textbf{x}}$};
		\node (y) at (-1.2,-0.8) {$\hat{\textbf{y}}$};
		\node (z) at (-2.2,-0.3) {$\hat{\textbf{z}}$};
    
		\draw [->, line width=0.4mm] (-2.0,-1.3) -- (-1.0,-1.3);
		\draw [->, line width=0.4mm] (-2.0,-1.3) -- (-2.0,-0.3);
		\draw [->, line width=0.4mm] (-2.0,-1.3) -- (-1.4,-0.8);
    
        \node[site](site1_1){$\uparrow\downarrow$};
        \node[ghost](site0_1)[left=of site1_1]{$\textbf{\dots}$};
        \node[site](site2_1)[right=of site1_1]{};
        \node[site](site3_1)[right=of site2_1]{$\downarrow$};
        \node[site](site4_1)[right=of site3_1]{};
        \node[site](site5_1)[right=of site4_1]{$\downarrow$};
        \node[site](site6_1)[right=of site5_1]{$\uparrow$};
        \node[site](site7_1)[right=of site6_1]{};
        \node[site](site8_1)[right=of site7_1]{$\uparrow$};
        \node[ghost](site9_1)[right=of site8_1]{$\textbf{\dots}$};
        
        \node[ghost](indx1)[below=of site1_1, yshift=0.2cm]{$U$};
        \node[ghost](indx2_3)[below=of site2_1, xshift=0.675cm, yshift=0.8cm]{$-t$};
        %\node[ghost](indx3_4)[below=of site3_1, xshift=0.675cm, yshift=0.2cm]{$-t$};
        \node[ghost](indx5_6)[below=of site5_1, xshift=0.675cm, yshift=0.2cm]{$V$};
        \node[ghost](indx8)[below=of site8_1, yshift=0.2cm]{$-\mu$};
        
        \node[box](Vbox)[right=of site4_1, xshift=-0.1cm]{};
        
        %\draw[->-, bend left = 40](site3_1.south) to (site2_1.south);
        %\draw[->-, bend right = 40](site3_1.south) to (site4_1.south);
        
        \draw[line width = 0.5mm, -](site0_1) to (site1_1);
        \draw[line width = 0.5mm, -](site1_1) to (site2_1);
        \draw[line width = 0.5mm, -](site2_1) to (site3_1);
        \draw[line width = 0.5mm, -](site3_1) to (site4_1);
        \draw[line width = 0.5mm, -](site4_1) to (site5_1);
        \draw[line width = 0.5mm, -](site5_1) to (site6_1);
        \draw[line width = 0.5mm, -](site6_1) to (site7_1);
        \draw[line width = 0.5mm, -](site7_1) to (site8_1);
        \draw[line width = 0.5mm, -](site8_1) to (site9_1);
        
        \node[site](site1_2)[above=of site1_1]{$\uparrow\downarrow$};
        \node[ghost](site0_2)[left=of site1_2]{$\textbf{\dots}$};
        \node[site](site2_2)[right=of site1_2]{};
        \node[site](site3_2)[right=of site2_2]{$\downarrow$};
        \node[site](site4_2)[right=of site3_2]{};
        \node[site](site5_2)[right=of site4_2]{$\downarrow$};
        \node[site](site6_2)[right=of site5_2]{$\uparrow$};
        \node[site](site7_2)[right=of site6_2]{};
        \node[site](site8_2)[right=of site7_2]{$\uparrow$};
        \node[ghost](site9_2)[right=of site8_2]{$\textbf{\dots}$};
        
        \draw[line width = 0.5mm, -](site0_2) to (site1_2);
        \draw[line width = 0.5mm, -](site1_2) to (site2_2);
        \draw[line width = 0.5mm, -](site2_2) to (site3_2);
        \draw[line width = 0.5mm, -](site3_2) to (site4_2);
        \draw[line width = 0.5mm, -](site4_2) to (site5_2);
        \draw[line width = 0.5mm, -](site5_2) to (site6_2);
        \draw[line width = 0.5mm, -](site6_2) to (site7_2);
        \draw[line width = 0.5mm, -](site7_2) to (site8_2);
        \draw[line width = 0.5mm, -](site8_2) to (site9_2);
        
        \node[ghost](tperp)[right=of site8_1, yshift=0.675cm, xshift=-0.8cm]{$t_\perp$};
        
        %\draw[<->, bend right = 30](site8_1.east) to (site8_2.east);
        
        \node[site](site1_3)[above=of site1_2]{$\uparrow\downarrow$};
        \node[ghost](site0_3)[left=of site1_3]{$\textbf{\dots}$};
        \node[site](site2_3)[right=of site1_3]{};
        \node[site](site3_3)[right=of site2_3]{$\downarrow$};
        \node[site](site4_3)[right=of site3_3]{};
        \node[site](site5_3)[right=of site4_3]{$\downarrow$};
        \node[site](site6_3)[right=of site5_3]{$\uparrow$};
        \node[site](site7_3)[right=of site6_3]{};
        \node[site](site8_3)[right=of site7_3]{$\uparrow$};
        \node[ghost](site9_3)[right=of site8_3]{$\textbf{\dots}$};
        
        \draw[line width = 0.5mm, -](site0_3) to (site1_3);
        \draw[line width = 0.5mm, -](site1_3) to (site2_3);
        \draw[line width = 0.5mm, -](site2_3) to (site3_3);
        \draw[line width = 0.5mm, -](site3_3) to (site4_3);
        \draw[line width = 0.5mm, -](site4_3) to (site5_3);
        \draw[line width = 0.5mm, -](site5_3) to (site6_3);
        \draw[line width = 0.5mm, -](site6_3) to (site7_3);
        \draw[line width = 0.5mm, -](site7_3) to (site8_3);
        \draw[line width = 0.5mm, -](site8_3) to (site9_3);
        
        \node[ghost](dotsLeft)[above=of site1_3, yshift = -0.3cm]{$\textbf{\vdots}$};
        \node[ghost](ddotsLeft)[right=of dotsLeft, xshift = -0.3cm, yshift = -0.2cm]{$\boldsymbol{\cdot^{\cdot^{\cdot}}}$};
        \node[ghost](dotsRight)[above=of site8_3, yshift = -0.3cm]{$\textbf{\vdots}$};
        \node[ghost](ddotsRight)[right=of dotsRight, xshift = -0.3cm, yshift = -0.2cm]{$\boldsymbol{\cdot^{\cdot^{\cdot}}}$};

        \draw[line width = 0.05mm, -](site1_1) to (site1_2);
        \draw[line width = 0.05mm, -](site2_1) to (site2_2);
        \draw[line width = 0.05mm, -](site3_1) to (site3_2);
        \draw[line width = 0.05mm, -](site4_1) to (site4_2);
        \draw[line width = 0.05mm, -](site5_1) to (site5_2);
        \draw[line width = 0.05mm, -](site6_1) to (site6_2);
        \draw[line width = 0.05mm, -](site7_1) to (site7_2);
        \draw[line width = 0.05mm, -](site8_1) to (site8_2);
        
        \draw[line width = 0.05mm, -](site1_3) to (site1_2);
        \draw[line width = 0.05mm, -](site2_3) to (site2_2);
        \draw[line width = 0.05mm, -](site3_3) to (site3_2);
        \draw[line width = 0.05mm, -](site4_3) to (site4_2);
        \draw[line width = 0.05mm, -](site5_3) to (site5_2);
        \draw[line width = 0.05mm, -](site6_3) to (site6_2);
        \draw[line width = 0.05mm, -](site7_3) to (site7_2);
        \draw[line width = 0.05mm, -](site8_3) to (site8_2);
    \end{scope}
\end{tikzpicture}
	}%
	{%
		\includegraphics{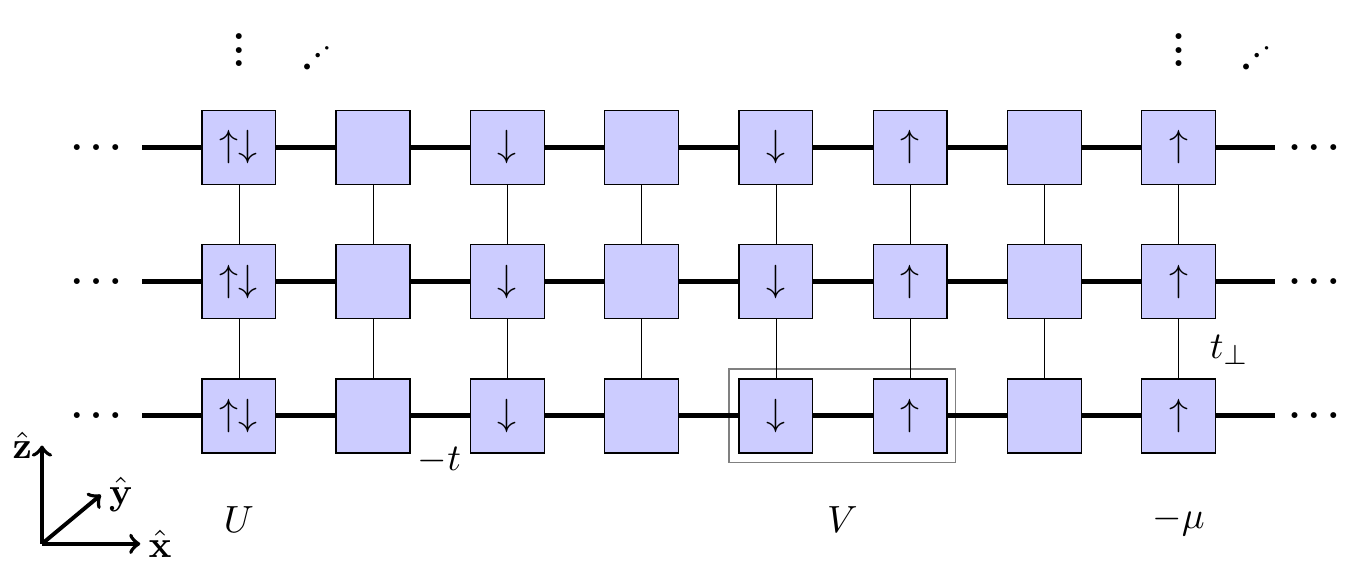}%
	}%
    \caption
    {
    	\label{fig:model}
    	Two dimensional schematic of the three dimensional model. 
    	For the sake of clarity, the extension of the system out of the plane is not shown here. 
    	Each box denotes a lattice site. 
    	The sites are coupled to chains in $\hat{\textbf{x}}$-direction, which is illustrated by the thick lines between the boxes. 
    	Furthermore, all chains are weakly coupled by the transverse hopping $t_\perp$. 
    	This way, we obtain an extension in $\hat{\textbf{y}}$ and $\hat{\textbf{z}}$-direction.
   	}
\end{figure}
As we aim to describe a \gls{3D} model system with a method that is mainly suitable for \gls{1D}, namely \gls{MPS}, we first need to identify a class of \gls{3D} models amenable to mapping onto an effective \gls{1D} description. 
Following the work of Bollmark et al.~\cite{Gunnar_coupled_chains, Gunnar_coupled_chains_electrons}, we focus on \gls{3D} systems constructed out of gapped \gls{1D} fermions. 
We arrange these \gls{1D} systems, which extend in the $\hat{\textbf{x}}$-direction, in parallel into a square array in the ${\hat{\textbf{y}}-\hat{\textbf{z}}}$-plane, forming effectively a cubic lattice. 
We choose fermion tunneling to be anisotropic in this lattice, denoted by $t_\perp$ in the $\hat{\textbf{y}}$- and $\hat{\textbf{z}}$-directions. 
Adapting from Bollmark et al.~\cite{Gunnar_coupled_chains_electrons}, we choose an extended Hubbard chain as the \gls{1D} building block. 
The Hamiltonian construct in this manner is illustrated in \cref{fig:model} and is given by
\begin{align}
    \hat{H} = \hat{H}_0 + t_\perp \hat{H}_\perp \;,
    \label{eq:Hamiltoian_short_form}
\end{align}
with 
\begin{align}
    \hat{H}_0 = &	- t \sum_{n=1}^{L-1} \sum_{\sigma\in\{\uparrow,\downarrow\}} \sum_{\{\textbf{R}_i\}} 
	    				\left( 
	    					\hat{c}^{\dagger}_{n+1,\textbf{R}_i,\sigma} \hat{c}^{\nodagger}_{n,\textbf{R}_i,\sigma} + \mathrm{h.c.}
	    				\right) 
    				- \mu \sum_{n=1}^L \sum_{\sigma\in\{ \uparrow,\downarrow\}} \sum_{\{\textbf{R}_i\}} \hat{n}_{n,\textbf{R}_i,\sigma} \\
    			&	+ U \sum_{n=1}^L \sum_{\{\textbf{R}_i\}}\hat{n}_{n,\textbf{R}_i,\uparrow} \hat{n}_{n,\textbf{R}_i,\downarrow} 
    				+ V \sum_{n=1}^{L-1} \sum_{\sigma,\sigma^{\prime}\in\{\uparrow,\downarrow\}} \sum_{\{\textbf{R}_i\}} \hat{n}_{n+1,\textbf{R}_i,\sigma} \hat{n}_{n,\textbf{R}_i,\sigma^{\prime}} \;,
    \label{eq:H0}
\end{align}
and
\begin{align}
    \hat{H}_\perp = - \sum_{n=1}^L \sum_{\sigma\in\{\uparrow,\downarrow\}} \sum_{\{\textbf{R}_i\}} \sum_{\hat{\textbf{a}}\in\{\hat{\textbf{y}},\hat{\textbf{z}}\}} 
    				  \left(
    				  	\hat{c}^{\dagger}_{n,\textbf{R}_i+\hat{\textbf{a}},\sigma} \hat{c}^{\nodagger}_{n,\textbf{R}_i,\sigma} + \mathrm{h.c.}
    				  \right) \;.
    \label{eq:Hperp}
\end{align}
Here, $\hat{c}^{\dagger}_{n,\textbf{R}_i,\sigma}$ and $\hat{c}^{\nodagger}_{n,\textbf{R}_i,\sigma}$ denote the fermionic creation and annihilation operators on site $n$ and for spin $\sigma$ on a chain that is labeled by the index $\textbf{R}_i$. 
They obey the anticommutation relations ${\{\hat{c}^{\nodagger}_i, \hat{c}^{\dagger}_j\} \equiv \hat{c}^{\nodagger}_i \hat{c}^{\dagger}_j + \hat{c}^{\dagger}_j \hat{c}^{\nodagger}_i = \delta_{ij}}$ and ${\{\hat{c}^{\nodagger}_i, \hat{c}^{\nodagger}_j\} = \{\hat{c}^{\dagger}_i,\hat{c}^{\dagger}_j\} = 0}$. 
The indices $i$ and $j$ stand for different combinations of $n,\textbf{R}_i$, and $\sigma$. 
The operator ${\hat{n}_{n,\textbf{R}_i,\sigma} = \hat{c}^{\dagger}_{n, \textbf{R}_i,\sigma} \hat{c}^{\nodagger}_{n, \textbf{R}_i,\sigma}}$ is the particle number operator for the corresponding site, chain, and spin. 
We use open boundary conditions and include a term for the chemical potential $\mu$. 
The latter allows us to control the number of particles in the system. 
The only non\hyp \gls{1D} term is the transverse hopping $\hat{H}_\perp$. 
We are able to eliminate the beyond\hyp \gls{1D} nature of this term through a combination of perturbation theory on the transverse hopping and a \gls{MF} decoupling of adjacent \gls{1D} systems. 
In the following we briefly recap the key steps, a detailed derivation of this approach can be found in the publications of Bollmark et al.~\cite{Gunnar_coupled_chains, Gunnar_coupled_chains_electrons}.
Since we are interested in a model system for \gls{SC}, we specify ${U<0}$ in the chain\hyp Hamiltonian \cref{eq:H0}. 
This negative\hyp $U$ term gives rise to pairing of opposite\hyp spin fermions already in isolated systems at ${t_\perp=0}$. 
This is expressed by the finite spin gap $\Delta E_s$ and a finite pairing energy $\Delta E_p$ of these isolated chains, defined as follows:
\begin{align}
   	\Delta E_s(N) &\equiv \mathcal{E}_0(1, N) - \mathcal{E}_0(0, N),\\
   	\Delta E_p(N) &\equiv 2\mathcal{E}_0\left(\frac{1}{2}, N+1\right) - \mathcal{E}_0(0, N) - \mathcal{E}_0(0, N+2) \;.
    \label{eq:energy_gaps}
\end{align}
Here, $\mathcal{E}_0(S_z, N)$ denotes the ground\hyp state energy of Hamiltonian $\hat{H}_0$ for a single chain\hyp index at total spin $S_z$ and total number of fermions $N$. 
Thus, $\Delta E_s$ and $\Delta E_p$ represent the minimal energy required for flipping a spin inside a chain and for breaking up a pair on a chain by moving one constituent to another chain in the full \gls{3D} system, respectively. 
From the definitions, it is easy to see that ${\Delta E_s \leq \Delta E_p}$, and for our specific choice of \gls{1D} systems ${\Delta E_s=\Delta E_p}$. 
As outlined in the following, $\Delta E_p$ becomes important in the actual numerical routine, directly entering the effective Hamiltonian \cref{eq:final_eff_MF_Ham}. 
In practice, we can determine $\Delta E_p$ from a single chain via an extrapolation in the system size ${L \rightarrow \infty}$.
To carry out the second\hyp order perturbation theory in $\hat{H}_\perp$ -- specifically in $t_\perp/\Delta E_p$ -- we follow \cite{cohen-tannoudji_atomPhotonInteractions}. 
We sort the eigenenergies $E_{i,\alpha}$ of $\hat{H}_0$, i.e., ${\hat{H}_0\ket{i,\alpha} = E_{i,\alpha}\ket{i,\alpha}}$, into a lowest\hyp energy manifold $E_{i,\alpha=0}$, where $i$ indexes the states within this manifold. 
In this manifold, there are no broken pairs. 
The high\hyp energy manifold $E_{i,\alpha=1}$ is at least $\Delta E_p$ above the low\hyp energy manifold, corresponding to excited states with at least one broken pair, i.e.,  where the pair\hyp constituents have moved onto separate chains. 
In the perturbative regime, we thus assume 
\begin{align}
    |E_{i,\alpha}-E_{j,\alpha}| \ll |E_{i,\alpha}-E_{j,\beta}|;\quad\alpha\ne\beta
    \label{eq:EnergySpectrumManifolds}
\end{align}
to hold.
We therefore target a small transverse hopping strength $t_\perp$ with respect to $\Delta E_s$ and $\Delta E_p$. 
Introducing the projector onto the lowest\hyp energy manifold ${\hat{P}_0=\sum_{i} | E_{i,0} \rangle\langle E_{i,0} |}$, the second\hyp order perturbation theory for Hamiltonian \cref{eq:Hamiltoian_short_form} yields:
\begin{align}
    \hat{H}^0_\mathrm{eff}=\hat{P}_0\hat{H}_0\hat{P}_0 - \frac{t_\perp^2}{\Delta E_p}\hat{P}_0\hat{H}_\perp^2\hat{P}_0 \;.
    \label{eq:H_eff0_afterPerturbationstuff}
\end{align}
Written explicitly, $\hat{H}_\perp^2$ is
\begin{align}
    \hat{H}_\perp^2 	=& \sum_{n,m=1}^L \sum_{\sigma\in\{\uparrow,\downarrow\}} \sum_{\{\textbf{R}_i\}} \sum_{\hat{\textbf{a}}\in\{\hat{\textbf{y}},\hat{\textbf{z}}\}} 
    						\left(
    							\hat{c}^{\dagger}_{n,\textbf{R}_i+\hat{\textbf{a}},\sigma} \hat{c}^{\nodagger}_{n,\textbf{R}_i,\sigma} \hat{c}^{\dagger}_{m,\textbf{R}_i+\hat{\textbf{a}},-\sigma} \hat{c}^{\nodagger}_{m,\textbf{R}_i,-\sigma} + \mathrm{h.c.} 
    						\right)\nonumber\\
    					 &+\sum_{n,m=1}^L \sum_{\sigma\in\{\uparrow,\downarrow\}} \sum_{\{\textbf{R}_i\}} \sum_{\hat{\textbf{a}}\in\{\hat{\textbf{y}},\hat{\textbf{z}}\}} 
    						\left(
    							\hat{c}^{\dagger}_{n,\textbf{R}_i+\hat{\textbf{a}},\sigma} \hat{c}^{\nodagger}_{n,\textbf{R}_i,\sigma} \hat{c}^{\dagger}_{m,\textbf{R}_i,\sigma} \hat{c}^{\nodagger}_{m,\textbf{R}_i+\hat{\textbf{a}},\sigma} + \mathrm{h.c.}
    						\right)\\
    					=& \hat{H}_\mathrm{pair} + \hat{H}_\mathrm{exc} \;.
    \label{eq:H_perp^2_spin_conserving}
\end{align}
Within \cref{eq:H_perp^2_spin_conserving}, we identify two contributions, namely a pairing term $\hat{H}_\mathrm{pair}$, which denotes the hopping of electron\hyp electron pairs of opposite spin between neighboring chains and an exchange term $\hat{H}_\mathrm{exc}$, denoting the exchange of particles of the same spin between neighboring chains.
In the following we use \gls{MF} theory to eliminate the non\hyp \gls{1D} nature of $\hat{H}_\perp^2$. 
Here, we make use of the relation 
\begin{align}
    c_i^{(\dagger)} c_j^{(\dagger)} = \left(c_i^{(\dagger)} c_j^{(\dagger)} - \braket{c_i^{(\dagger)} c_j^{(\dagger)}}\right) + \braket{c_i^{(\dagger)} c_j^{(\dagger)}} \;,
    \label{eq:mf-ansatz}
\end{align}
and assume ${\left(c_i^{(\dagger)} c_j^{(\dagger)} - \braket{c_i^{(\dagger)} c_j^{(\dagger)}}\right)}$ to be small. 
We, moreover, assume 
\begin{align}
	\braket{\hat{c}^{\nodagger}_{n,\uparrow} \hat{c}^{\nodagger}_{m,\downarrow}} = \braket{\hat{c}^{\nodagger}_{n,\textbf{R}_i,\uparrow} \hat{c}^{\nodagger}_{m,\textbf{R}_i,\downarrow}} = \braket{\hat{c}^{\nodagger}_{n,\textbf{R}_i+\hat{\textbf{a}},\uparrow} \hat{c}^{\nodagger}_{m,\textbf{R}_i+\hat{\textbf{a}},\downarrow}} \;,
	\label{eq:identical_order_parms}
\end{align} 
which means that all the chains are exact copies of each other. 
We end up with an effectively \gls{1D} expression for a Hamiltonian describing a multidimensional model, namely
\begin{align}
    \hat{H}_\mathrm{eff}^\mathrm{MF} =& \hat{H}_0 - \sum_{n,m=1}^L \left( \alpha_{n,m}^*\hat{c}^{\nodagger}_{n,\uparrow} \hat{c}^{\nodagger}_{m,\downarrow} + \alpha_{n,m} \hat{c}^{\dagger}_{m,\downarrow} \hat{c}^{\dagger}_{n,\uparrow} \right)\nonumber\\
    								 & + \sum_{n=1}^L \sum_{\sigma\in\{\uparrow,\downarrow\}} \sum_{r=1}^{L-n} \left(\beta_{n,r,\sigma}^{*} \hat{c}^{\dagger}_{n+r,\sigma} \hat{c}^{\nodagger}_{n,\sigma} + \beta^{\nodagger}_{n,r,\sigma} \hat{c}^{\dagger}_{n,\sigma} \hat{c}^{\nodagger}_{n+r,\sigma}\right)
\label{eq:final_eff_MF_Ham}
\end{align}
with
\begin{align}
    \alpha_{n,m} &= \frac{2z_ct_\perp^2}{\Delta E_p} \braket{\hat{c}^{\nodagger}_{n,\uparrow}\hat{c}^{\nodagger}_{m,\downarrow}} \quad \mathrm{and} 
    \label{eq:self-cons-param-pairing}\\
    \beta_{n,r,\sigma} &= \frac{2z_ct_\perp^2}{\Delta E_p}v\braket{\hat{c}^{\dagger}_{n+r,\sigma} \hat{c}^{\nodagger}_{n,\sigma}} \;,
    \label{eq:self-cons-param-exchange}
\end{align}
and thus identify $\alpha_{n,m}$ with the \gls{MF}\hyp approximated pairing part of \cref{eq:H_perp^2_spin_conserving} and $\beta_{n,r,\sigma}$ with its exchange part.
Here, we introduced the coordination number $z_c$, which denotes the number of neighboring chains. 
In our case $z_c = 4$, as the chains are assembled into a \gls{2D} square grid in the $\hat{\textbf{y}}-\hat{\textbf{z}}$-plane. 
The parameters $\alpha_{n,m}$ and $\beta_{n,r,\sigma}$ are the so\hyp called \gls{MF} parameters, meaning they need to be calculated self\hyp consistently for all times. 
The work in \cite{Gunnar_coupled_chains, Gunnar_coupled_chains_electrons} explains this for the ground state and for the finite\hyp temperature equilibrium of the \gls{3D} system. 
There, the authors demonstrate that the \gls{MPS}+\gls{MF} approach for equilibrium systems produces the correct physics compared against \gls{QMC}, in regimes in which the latter approach is quasi\hyp exact, in a negative\hyp $U$ Hubbard model on a \gls{2D} square lattice with anisotropic tunneling. 
That work also shows that the error in $T_c$ for the \gls{SC} state due to the \gls{MF} approximation within the \gls{MPS}+\gls{MF} framework is a quasi\hyp constant one in $t_\perp$ over a significant range. 
Moreover, at zero temperature, the overestimation due to the \gls{SC} order parameter becomes systematically better as $t_\perp$ decreases.
Based on the good performance of the \gls{MPS}+\gls{MF} scheme in equilibrium, the present work is concerned with the performance of the self\hyp consistent evaluation of the \gls{MF} parameters \cref{eq:self-cons-param-pairing} and \cref{eq:self-cons-param-exchange} for a time\hyp evolving system. 
Since the present work aims to test and benchmark the method itself, in the following we are working with the simplest possible version of the Hamiltonian \cref{eq:final_eff_MF_Ham}. 
We neglect the exchange term $\beta_{n,r,\sigma}$ and allow only for site\hyp independent onsite pairing, meaning ${\alpha_{n,m} \equiv \alpha_{n,n} \equiv \alpha}$. 
This leads to
\begin{align}
    \hat{H}_\mathrm{eff}^\mathrm{MF} = \hat{H}_0 - \sum_{n}\left(\alpha^{*} \hat{c}^{\nodagger}_{n,\uparrow} \hat{c}^{\nodagger}_{n,\downarrow} + \alpha \hat{c}^{\dagger}_{n,\downarrow} \hat{c}^{\dagger}_{n,\uparrow} \right)
    \label{eq:final_eff_MF_Ham_simplified}
\end{align}
with
\begin{align}
    \alpha = \frac{1}{L} \frac{2z_ct_\perp^2}{\Delta E_p} \sum_{n=1}^L \braket{\hat{c}^{\nodagger}_{n,\uparrow} \hat{c}^{\nodagger}_{n,\downarrow}} \;.
    \label{eq:self-cons-param-pairing_simplified}
\end{align}
In this last expression we are adapting the evaluation of the order parameter $\alpha$ to the open boundary conditions. 
Obtaining $\alpha$ from an average across the entire system removes the spatial variation that is solely due to these open boundaries.
\section{\label{sec:algorithm}MPS+MF-Algorithm for self-consistent time-evolution}
The expectation values needed to compute the \gls{MF} parameter $\alpha$ in \cref{eq:self-cons-param-pairing_simplified} are computed using a self\hyp consistent scheme
for both the time\hyp evolution and for the ground\hyp state search of our model system. 
In this section a schematic description of the time\hyp evolution routine is presented, which is one of our main results.
The algorithm is based on the work of H. Strand et al.~published in \cite{Hugo_Strand_Nonequilibrium-dynamics_BoseHubbard}, where a non\hyp equilibrium version of real\hyp time \gls{DMFT} for bosons is introduced. 
Our work incorporates this real\hyp time scheme into a \gls{MPS} framework and adapts it to \gls{3D} lattices of correlated fermions built from weakly coupled \gls{1D} systems.
All results obtained in the following were generated with Ian McCulloch's matrix product toolkit \cite{mps_toolkit_Ian}. 
The initial ground states from which the time evolution proceeds were generated from a self\hyp consistent scheme introduced by Bollmark et al.~in \cite{Gunnar_coupled_chains}, which is also briefly described in \cref{sec:appendix_self-cons-GS}.
%
%

%\subsection
%{
	\label{sec:self-cons-tevol_algorithm}
%	Self\hyp consistent time evolution
%}
%
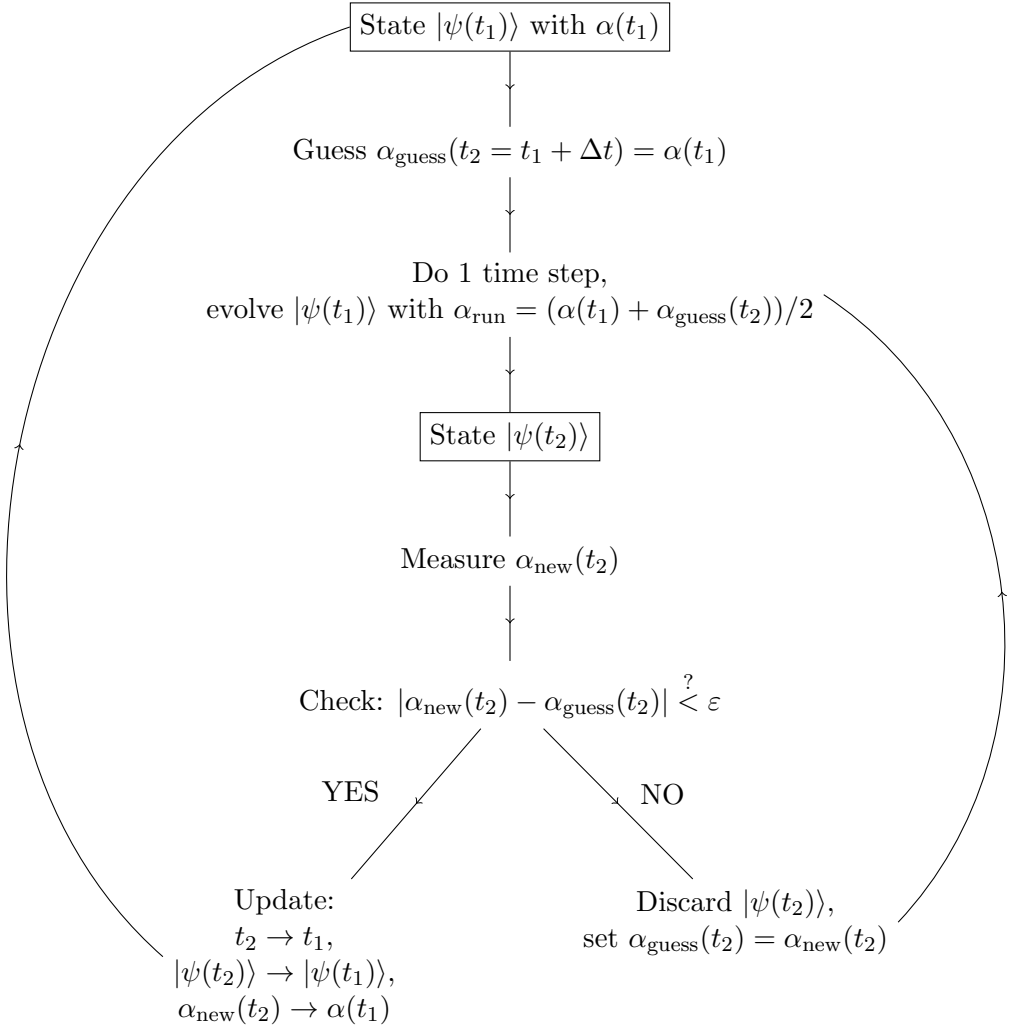
\begin{figure}[!t]
	\centering
	\ifthenelse{\boolean{buildtikzpics}}%
	{%
		%\tikzset{external/export next=false}
\tikzsetnextfilename{self-cons-tevol_scheme}
\begin{tikzpicture}[
box/.style={rectangle, draw=black, align=center},
ghost/.style={align=center}
]
	\clip (-6.75,0.5) rectangle (6.75,-13.3);
    \begin{scope}[node distance=1.0cm]
        \node[box](iniState){State $\ket{\psi(t_1)}$ with $\alpha(t_1)$};
        \node[ghost](step0)[below=of iniState]{Guess $\alpha_{\text{guess}}(t_2 = t_1+\Delta t) = \alpha(t_1)$};
        \node[ghost](step1)[below=of step0]{Do 1 time step,\\ evolve $\ket{\psi(t_1)}$ with $\alpha_{\text{run}}=(\alpha(t_1)+\alpha_\text{guess}(t_2))/2$};
        \node[box](newState)[below=of step1]{State $\ket{\psi(t_2)}$};
        \node[ghost](step2)[below=of newState]{Measure $\alpha_\text{new}(t_2)$};
        \node[ghost](step3)[below=of step2]{Check: $\left|\alpha_\text{new}(t_2)-\alpha_\text{guess}(t_2)\right|\stackrel{?}{<}\varepsilon$};
        %\node[ghost](yes)[below=of step3, xshift=-3cm]{YES};
        %\node[ghost](no)[below=of step3, xshift=3cm]{NO};
        \node[ghost](stepYes0)[below=of step3, xshift=-3cm, yshift = -1.0cm]{Update:\\ $t_2\rightarrow t_1$,\\ $\ket{\psi(t_2)}\rightarrow\ket{\psi(t_1)}$,\\ $\alpha_\text{new}(t_2)\rightarrow\alpha(t_1)$};
        \node[ghost](stepNo0)[below=of step3, xshift=3cm, yshift = -1.0cm]{Discard $\ket{\psi(t_2)}$,\\ set $\alpha_\text{guess}(t_2)=\alpha_\text{new}(t_2)$};
        
        \draw[->-, bend left=60](stepYes0.west) to (iniState.west);
        \draw[->-, bend right=50](stepNo0.east) to (step1.east);
        \draw[->-](iniState) to (step0);
        \draw[->-](step0) to (step1);
        \draw[->-](step1) to (newState);
        \draw[->-](newState) to (step2);
        \draw[->-](step2) to (step3);
        \draw[->-](step3) to (stepYes0) node[left, xshift=1.4cm, yshift=2.2cm]{YES};
        \draw[->-](step3) to (stepNo0) node[right, xshift=-1.4cm, yshift=1.7cm]{NO};
    \end{scope}
\end{tikzpicture}
	}%
	{%
		\includegraphics{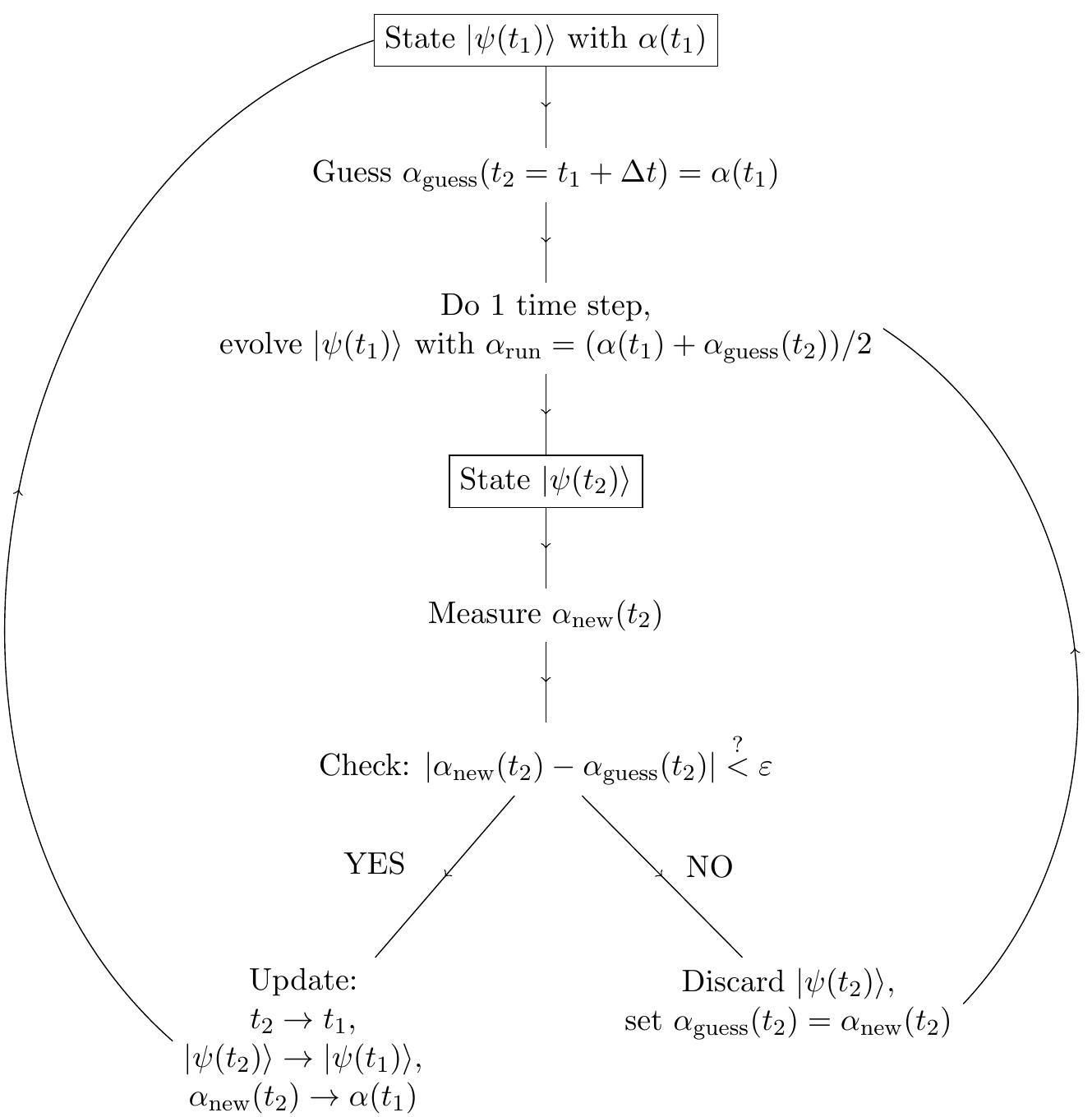}%
	}%
    \caption
    {
    	\label{fig:self_cons_tevol}
    	%´
    	Self consistency loop for one time step.
    	As the MF-parameter $\alpha$ depends on the state itself, a continuous adjustment of it is required.
    }
\end{figure}
At the beginning of each time step, we start with a state $\ket{\psi(t_1)}$ at time $t_1$, which we already have obtained before (either as a previous step or as initial state).
From this state, we measure the value of the \gls{MF} parameter $\alpha(t_1)$. 
Now, we guess which value $\alpha$ might take after one discrete time step $\mathrm{d}t$. 
In this work, at the start of the self\hyp consistency iterations for each time step, we just assume that the $\alpha$ value does not change at all. 
In any case, the guess for $\alpha$ at ${t_2= t_1 + \mathrm{d}t}$, is labeled $\alpha_\mathrm{guess}(t_2)$. 
Then, we evolve the system from $t_1$ to $t_2$ using the mean of $\alpha(t_1)$ and $\alpha_\mathrm{guess}$. 
From the resulting tentative $\ket{\psi(t_2)}$ we can once again measure the \gls{MF} parameter $\alpha_\mathrm{new}(t_2)$. 
Next we calculate the distance between the measured and the guessed value and compare it to a chosen precision $\varepsilon$,
\begin{align}
    \left|\alpha_\mathrm{new}(t_2)-\alpha_\mathrm{guess}(t_2)\right|<\varepsilon\quad\mathrm{with}\quad\varepsilon\ll1 \;.
    \label{eq:condition_alpha_timestep}
\end{align}
If \cref{eq:condition_alpha_timestep} is fulfilled, we keep the state $\ket{\psi(t_2)}$ and proceed with the next time step. 
Otherwise, we discard $\ket{\psi(t_2)}$ and repeat the time step using the mean of $\alpha(t_1)$ and $\alpha_\mathrm{new}(t_2)$. 
The loop is repeated until \cref{eq:condition_alpha_timestep} is fulfilled. 
A schematic of the algorithm is depicted in \cref{fig:self_cons_tevol}.
\section{\label{sec:resultsAndDiscussion}Transient \gls{SC} after a fast ramp of the nearest\hyp neighbor interaction}
In this section, we present our results using the self\hyp consistent \gls{MPS}+\gls{MF} scheme and find that in the extended Hubbard model \cref{eq:Hamiltoian_short_form} the \gls{BCS} order parameter for \gls{SC} grows in time and begins to oscillate around a finite value on the treated time scales.
This indicates the formation of transient \gls{SC}, which is the second main result of this paper.
In the following, all parameters are measured in units of the hopping parameter ${t \equiv 1}$. 
More specifically, we follow Paeckel et al.~\cite{Paeckel_superconductivity} and tune the system's parameters from a \gls{CDW} phase into a \gls{SC} phase.
However, we find that the sudden quench performed in \cite{Paeckel_superconductivity} is numerically less stable within the self\hyp consistent scheme (see \cref{sec:appendix_rampTime}), so we instead perform a fast ramp. 
In order to check the equilibrium phases of the \gls{3D} model we use the self\hyp consistent \gls{MPS}+\gls{MF} approach to compute the ground states using the routine introduced by Bollmark et al.~\cite{Gunnar_coupled_chains} for different parameters and measure the expectation value of the \gls{MF} parameter $\alpha$. 
We find that for ${t_\perp = 0.2}$, ${U = -4}$ and ${V = 0.25}$ the system possesses the main properties of a \gls{CDW} phase relevant for us, i.e., we find alternating occupation of the lattice sites by the electrons and a vanishing value of $\alpha$. 
For ${U = -4}$ and ${V = -0.25}$ instead, the system is \gls{SC}, as here ${\alpha \sim 10^{-1}}$ becomes finite and density oscillations less pronounced. 
These are the same parameters treated by Paeckel et al.~in \cite{Paeckel_superconductivity} for the purely \gls{1D} system.
Hence, we perform a fast ramp by tuning the values of the nearest\hyp neighbor interaction from ${V = 0.25}$ to ${V = -0.25}$ as further detailed below.
Since the effective Hamiltonian \cref{eq:final_eff_MF_Ham_simplified} depends on the \gls{MF} parameter $\alpha(t)$ the question of how to choose $\alpha_\mathrm{ini}:=\alpha(t=0)$ arises. 
For the \gls{CDW} system $\alpha=0$ and it is hence difficult for it to grow with the method outlined  in~\cref{fig:self_cons_tevol}.
Because of this, unless otherwise noted, our default value for this work is $\alpha_\mathrm{ini}=10^{-4}/\mathrm{d}t$, where $\mathrm{d}t$ is the size of the discretized time step of the simulation. 
Such a small yet finite value is justified by the fact that any system will either have a microscopic fraction of pairs in the center\hyp of\hyp mass zero\hyp momentum state to begin with, or such a fraction is generated during the ramp or quench. 
Scaling $\alpha_\mathrm{ini}$ inversely in $\mathrm{d}t$ ensures that simulations with different $\mathrm{d}t$ agree over long times, see \cref{fig:alphaEandRho_tevol_L30_chi250}.
The \gls{MF} term of the Hamiltonian causes the effective model to be no longer particle\hyp number conserving, hence, we need to adjust the value of the chemical potential $\mu$ corresponding to the system size and to the onsite repulsion $U$ in order to fix the average density of the total system.
From the ground\hyp state calculations we find the values of $\mu$ that are listed in \cref{tab:chemical_potential_U-4.0}. 
We keep the values of $\mu$, determined in this manner, fixed throughout the whole time evolution in order to keep our algorithm simple and stable. 
However, we still need to keep track of the overall density of our system during the time evolution to check if this assumption of a time\hyp independent chemical potential is justified. 
Indeed, for our simulations, the value of the density is preserved to a good accuracy over the time scales treated by us (see \cref{fig:alphaEandRho_tevol_L30_chi250,fig:alphaEandRho_tevol_Lvary_chi500}).
In general, however, it might be necessary to also include a variation of $\mu$ into the self\hyp consistency scheme. 
\begin{table}[!t]
    \centering
    \caption
    {
	    \label{tab:chemical_potential_U-4.0}
    	List of values for the chemical potential $\mu$ to obtain half filling for ${U = -4.0}$ and ${V = \pm 0.25}$ for various system sizes $L$.
    }
    \begin{tabular}{lccc} 
        \toprule
        $L$ 			& 12 	& 20 	& 30 	\\
        \midrule
	    $\mu(V=-0.25)$  & -2.44 & -2.47 & -2.48	\\
    	$\mu(V=0.25)$  	& -1.66 & -1.63 & -1.62 \\
    	\bottomrule
    \end{tabular}
\end{table}
\subsection{\label{sec:observablesInTime}Time evolution of the \acrshort{BCS} order parameter and of the total energy}
\begin{figure}[!t]
    \centering
	\ifthenelse{\boolean{buildtikzpics}}%
	{%
%		for i in L_*/U_*/V_-0.25/chi_500/dt_0.01/alphaIni_0.0001/Data_tevol.txt; do L=$(echo $i | pcregrep -o1 "^L_(.*)/U"); echo -n "$L "; cat $i | awk '{print($1,sqrt($2*$2+$3*$3))}' | sort -k 2 -g -r | head -n1; done
%		\tikzset{external/export next=false}
		\tikzsetnextfilename{Lvary_tRamp3p0_tend50_chi500_dt0p01_alphaIni0p0001}
		\begin{tikzpicture}
			\begin{groupplot}%
			[%
				group style =%
				{%
					group size			=	2 by 2,%
					vertical sep		=	0.25em,%
					horizontal sep		=	4.5em,%
					x descriptions at	=	edge bottom,%
				},%
				height		=	0.2\textheight,%
				width		=	0.5\textwidth-3.33pt,%
				xlabel		=	{time $t$},%
				xticklabels	=	{$0$, $3$, $20$, $40$},%
				xtick		=	{0,3,20,40},%
			]%
				\nextgroupplot%
				[%
					ylabel				=	{$\lvert \alpha \rvert\vphantom{\rho^{-5}}$},%
					yticklabels			=	{$\hphantom{-0.}0$, $0.02$, $0.04$, $0.06$, $0.08$},%
					ytick				=	{0, 0.02, 0.04, 0.06, 0.08},%
					scaled y ticks 		=	false,%
					ylabel style 		=	{yshift = -0.2em},%
				]%
					\coordinate (l1) at (rel axis cs:-0.25,0.85);%
					\addplot%
					[%
						color	=	colorA,%
						thick,%
					]%
						table%
						[%
							x expr = \thisrowno{0},%
							y expr = sqrt(\thisrowno{1}*\thisrowno{1}+\thisrowno{2}*\thisrowno{2})%
						]%
							{data/tRamp_3.0/tend_50/tperp_0.2/L_12/U_-4.0/V_-0.25/chi_500/dt_0.01/alphaIni_0.0001/Data_tevol.txt};%
					\addplot%
					[%
						color	=	colorB,%
						densely dashed,
						thick,%
					]%
						table%
						[%
							x expr = \thisrowno{0},%
							y expr = sqrt(\thisrowno{1}*\thisrowno{1}+\thisrowno{2}*\thisrowno{2})%
						]%
							{data/tRamp_3.0/tend_50/tperp_0.2/L_16/U_-4.0/V_-0.25/chi_500/dt_0.01/alphaIni_0.0001/Data_tevol.txt};%
					\addplot%
					[%
						color	=	colorC,%
						densely dotted,
						thick,%
					]%
						table%
						[%
							x expr = \thisrowno{0},%
							y expr = sqrt(\thisrowno{1}*\thisrowno{1}+\thisrowno{2}*\thisrowno{2})%
						]%
							{data/tRamp_3.0/tend_50/tperp_0.2/L_20/U_-4.0/V_-0.25/chi_500/dt_0.01/alphaIni_0.0001/Data_tevol.txt};%
					\addplot%
					[%
						color	=	colorD,%
						densely dashdotted,
						thick,%
					]%
						table%
						[%
							x expr = \thisrowno{0},%
							y expr = sqrt(\thisrowno{1}*\thisrowno{1}+\thisrowno{2}*\thisrowno{2})%
						]%
							{data/tRamp_3.0/tend_50/tperp_0.2/L_30/U_-4.0/V_-0.25/chi_500/dt_0.01/alphaIni_0.0001/Data_tevol.txt};%
					\coordinate (inset) at (axis description cs:0.175,0.95);%
				\nextgroupplot%
				[%
					ylabel			=	{$E/L\vphantom{\rho^{-5}}$},%
					ylabel style	=	{yshift = -0.2em},%
					legend pos			=	north east,%
					legend cell align	=	{left},%
					legend style =%
					{%
						draw	=	none,%
						fill	=	none,%
					},%
				]%
					\coordinate (l2) at (rel axis cs:-0.25,0.85);%
					\addplot%
					[%
						color	=	colorA,%
						thick,%
					]%
						table%
						[%
							x expr = \thisrowno{0},%
							y expr = \thisrowno{3}/12.0%
						]%
							{data/tRamp_3.0/tend_50/tperp_0.2/L_12/U_-4.0/V_-0.25/chi_500/dt_0.01/alphaIni_0.0001/Data_tevol.txt};%
					\addlegendentry{$L=12$};%
					\addplot%
					[%
						color	=	colorB,%
						densely dashed,%
						thick,%
					]%
						table%
						[%
							x expr = \thisrowno{0},%
							y expr = \thisrowno{3}/16.0%
						]%
							{data/tRamp_3.0/tend_50/tperp_0.2/L_16/U_-4.0/V_-0.25/chi_500/dt_0.01/alphaIni_0.0001/Data_tevol.txt};%
					\addlegendentry{$L=16$};%
					\addplot%
					[%
						color	=	colorC,%
						densely dotted,%
						thick,%
					]%
						table%
						[%
							x expr = \thisrowno{0},%
							y expr = \thisrowno{3}/20.0%
						]%
							{data/tRamp_3.0/tend_50/tperp_0.2/L_20/U_-4.0/V_-0.25/chi_250/dt_0.01/alphaIni_0.0001/Data_tevol.txt};%
					\addlegendentry{$L=20$};%
					\addplot%
					[%
						color	=	colorD,%
						densely dashdotted,%
						thick,%
					]%
						table%
						[%
							x expr = \thisrowno{0},%
							y expr = \thisrowno{3}/30.0%
						]%
							{data/tRamp_3.0/tend_50/tperp_0.2/L_30/U_-4.0/V_-0.25/chi_250/dt_0.01/alphaIni_0.0001/Data_tevol.txt};%
					\addlegendentry{$L=30$};%
				\nextgroupplot%
				[%
					ylabel			=	{$\varphi\left( \alpha \right) /\pi\vphantom{\rho^{-5}}$},%
					ylabel style	=	{yshift = -0.2em},%
					yticklabels		=	{$-1$, $-0.8$, $-0.6$, $-0.4$, $-0.2$, $\hphantom{-0.}0$, $1$},%
					ytick			=	{-1, -0.8, -0.6, -0.4, -0.2, 0, 1},%
				]%
					\coordinate (l3) at (rel axis cs:-0.25,0.85);%
					\addplot%
					[%
						color	=	colorA,%
						thick,%
					]%
						table%
						[%
							x expr = \thisrowno{0},%
							y expr = %
							(%
								\thisrowno{1} > 0 ? %
								(%
									rad(atan(\thisrowno{2}/\thisrowno{1}))%
								)%
								:%
								(%
									\thisrowno{2} > 0 ?%
									(%
										(180+atan(\thisrowno{2}/\thisrowno{1}))%
									)%
									:%
									(%
										rad(-180+atan(\thisrowno{2}/\thisrowno{1}))%
									)%
								)%
							)/(pi),%
						]%
							{data/tRamp_3.0/tend_50/tperp_0.2/L_12/U_-4.0/V_-0.25/chi_500/dt_0.01/alphaIni_0.0001/Data_tevol.txt};%
					\addplot%
					[%
						color	=	colorB,%
						densely dashed,%
						thick,%
					]%
						table%
						[%
							x expr = \thisrowno{0},%
							y expr = %
							(%
								\thisrowno{1} > 0 ? %
								(%
									rad(atan(\thisrowno{2}/\thisrowno{1}))%
								)%
								:%
								(%
									\thisrowno{2} > 0 ?%
									(%
										(180+atan(\thisrowno{2}/\thisrowno{1}))%
									)%
									:%
									(%
										rad(-180+atan(\thisrowno{2}/\thisrowno{1}))%
									)%
								)%
							)/(pi),%
						]%
							{data/tRamp_3.0/tend_50/tperp_0.2/L_16/U_-4.0/V_-0.25/chi_500/dt_0.01/alphaIni_0.0001/Data_tevol.txt};%
					\addplot%
					[%
						color	=	colorC,%
						densely dotted,%
						thick,%
					]%
						table%
						[%
							x expr = \thisrowno{0},%
							y expr = %
							(%
								\thisrowno{1} > 0 ? %
								(%
									rad(atan(\thisrowno{2}/\thisrowno{1}))%
								)%
								:%
								(%
									\thisrowno{2} > 0 ?%
									(%
										rad(180+atan(\thisrowno{2}/\thisrowno{1}))%
									)%
									:%
									(%
										rad(-180+atan(\thisrowno{2}/\thisrowno{1}))%
									)%
								)%
							)/(pi),%
						]%
							{data/tRamp_3.0/tend_50/tperp_0.2/L_20/U_-4.0/V_-0.25/chi_500/dt_0.01/alphaIni_0.0001/Data_tevol.txt};%
					\addplot%
					[%
						color	=	colorD,%
						densely dashdotted,%
						thick,%
					]%
						table%
						[%
							x expr = \thisrowno{0},%
							y expr = %
							(%
								\thisrowno{1} > 0 ? %
								(%
									rad(atan(\thisrowno{2}/\thisrowno{1}))%
								)%
								:%
								(%
									\thisrowno{2} > 0 ?%
									(%
										rad(180+atan(\thisrowno{2}/\thisrowno{1}))%
									)%
									:%
									(%
										rad(-180+atan(\thisrowno{2}/\thisrowno{1}))%
									)%
								)%
							)/(pi),%
						]%
							{data/tRamp_3.0/tend_50/tperp_0.2/L_30/U_-4.0/V_-0.25/chi_500/dt_0.01/alphaIni_0.0001/Data_tevol.txt};%
				\nextgroupplot%
				[%
					ylabel			=	{$\left( 1-\rho \right)/10^{-5}$},%
					yticklabels		=	{$\hphantom{-0.}0$, $0.5$, $1$, $2$, $3$},%
					ytick			=	{0, 0.5, 1, 2, 3},%
					ylabel style	=	{yshift = -0.2em},%
				]%
					\coordinate (l4) at (rel axis cs:-0.25,0.85);%
					\addplot%
					[%
						color	=	colorA,%
						thick,%
					]%
						table%
						[%
							x expr = \thisrowno{0},%
							y expr = \eval{10^5-\thisrowno{15}*10^5}%
						]%
							{data/tRamp_3.0/tend_50/tperp_0.2/L_12/U_-4.0/V_-0.25/chi_500/dt_0.01/alphaIni_0.0001/Data_tevol.txt};%
					\addplot%
					[%
						color	=	colorB,%
						densely dashed,%
						thick,%
					]%
						table%
						[%
							x expr = \thisrowno{0},%
							y expr = \eval{10^5-\thisrowno{15}*10^5}%
						]%
							{data/tRamp_3.0/tend_50/tperp_0.2/L_16/U_-4.0/V_-0.25/chi_500/dt_0.01/alphaIni_0.0001/Data_tevol.txt};%
					\addplot%
					[%
						color	=	colorC,%
						densely dotted,%
						thick,%
					]%
						table%
						[%
							x expr = \thisrowno{0},%
							y expr = \eval{10^5-\thisrowno{15}*10^5}%
						]%
							{data/tRamp_3.0/tend_50/tperp_0.2/L_20/U_-4.0/V_-0.25/chi_500/dt_0.01/alphaIni_0.0001/Data_tevol.txt};%
					\addplot%
					[%
						color	=	colorD,%
						densely dashdotted,%
						thick,%
					]%
						table%
						[%
							x expr = \thisrowno{0},%
							y expr = \eval{10^5-\thisrowno{15}*10^5}%
						]%
							{data/tRamp_3.0/tend_50/tperp_0.2/L_30/U_-4.0/V_-0.25/chi_500/dt_0.01/alphaIni_0.0001/Data_tevol.txt};%
			\end{groupplot}%
			\begin{axis}%
			[%
				at				=	(inset),%
				anchor			=	north west,%
				xlabel			=	{\footnotesize $1/L$},%
				ylabel			=	{\footnotesize $t_{\mathrm{SC}}$},%
				width			=	0.275\textwidth,%
				height			=	0.125\textheight,%
				xmin			=	0,
				xmax			=	0.1,
				ymin			=	38,
				ymax			=	54,
				yticklabels		=	{\footnotesize $40$, \footnotesize $45$, \footnotesize $50$},%
				ytick			=	{40,45,50},%
				ylabel style	=	{yshift = -0.4em},%
				xticklabels		=	{\footnotesize $0$, \footnotesize $0.05$, \footnotesize $0.1$},%
				xtick			=	{0, 0.05, 0.1},%
				xlabel style	=	{yshift = 0.2em},%
				scaled y ticks 	=	false,%
			]%
				\addplot%
				[%
					color	=	black,%
					mark	=	o,
					only marks,
					thick,%
				]%
					table%
					[%
						x expr = 1.0/\thisrowno{0},%
						y expr = \thisrowno{1}%
					]%
						{data/tRamp_3.0/tend_50/tperp_0.2/localMaxima.dat};%
				\addplot%
				[%
					color	=	gray,%
					domain	=	0:0.1,
					thick,%
				]%
					{linearFct(-135.596, 52.856)};
				\addplot%
				[%
					color	=	blue,%
					domain	=	0:0.1,
					thick,%
				]%
					{quadFct(-1416.78, 30.6372, 48.4545)};
			\end{axis}%
			\node at (l1) {\subfloat[\label{fig:AbsAlpha_tevol_Lvary_chi500}]{}};%
			\node at (l3) {\subfloat[\label{fig:PhaseAlpha_tevol_Lvary_chi500}]{}};%
			\node at (l2) {\subfloat[\label{fig:E_tevol_Lvary_chi500}]{}};%
			\node at (l4) {\subfloat[\label{fig:Rho_tevol_Lvary_chi500}]{}};%
		\end{tikzpicture}
	}%
	{%
		\includegraphics{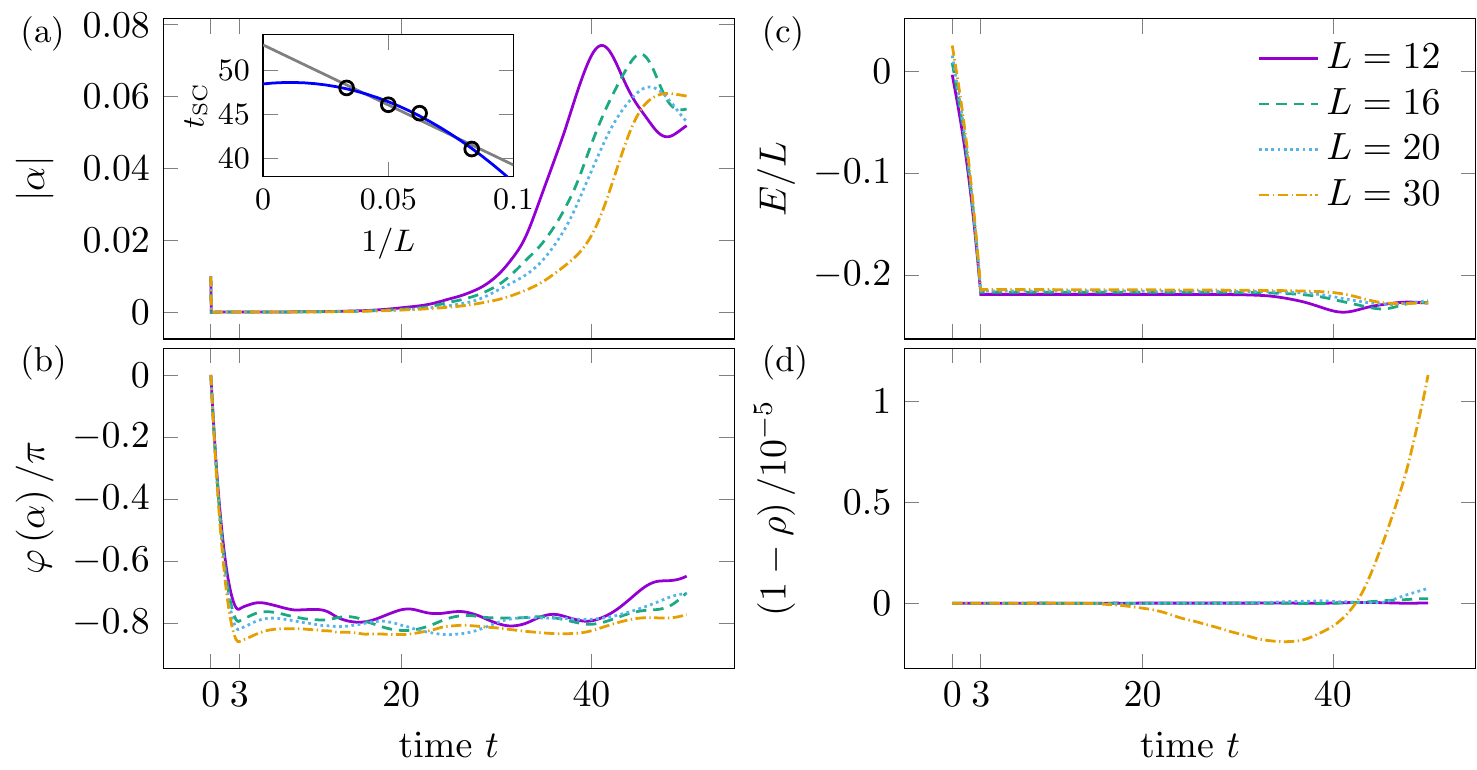}
		
		{\phantomsubcaption\label{fig:AbsAlpha_tevol_Lvary_chi500}}%
		{\phantomsubcaption\label{fig:PhaseAlpha_tevol_Lvary_chi500}}%
		{\phantomsubcaption\label{fig:E_tevol_Lvary_chi500}}%
		{\phantomsubcaption\label{fig:Rho_tevol_Lvary_chi500}}%
	}%
    \caption
    {
	    \label{fig:alphaEandRho_tevol_Lvary_chi500}
    	Evolution of the considered parameters in time during and after a ramp on various system sizes. 
		Magnitude \protect\subfigref{fig:AbsAlpha_tevol_Lvary_chi500} and phase \protect\subfigref{fig:PhaseAlpha_tevol_Lvary_chi500} of the dynamics of the \gls{MF} parameter $\alpha$.
    	The inset in \protect\subfigref{fig:AbsAlpha_tevol_Lvary_chi500} shows the time at which the first local maximum in $\left|\alpha\right|$ occurs plotted against the inverse of the system size $1/L$. 
    	Both, the linear and the quadratic fit suggest a finite and comparable value in the limit $L\rightarrow\infty$.
    	Evolution of the total energy of the system \protect\subfigref{fig:E_tevol_Lvary_chi500} and the total density \protect\subfigref{fig:Rho_tevol_Lvary_chi500}. 
    	All data were obtained with a bond dimension of ${\chi=500}$, an initial guess of the \gls{MF} parameter of ${\alpha_\mathrm{ini} = 10^{-4}/\mathrm{d}t}$, a ramp time window $\Delta t_{\mathrm{ramp}}=3$, and the chemical potential was taken from \cref{tab:chemical_potential_U-4.0}. 
    }
\end{figure}

\begin{figure}[!t]
    \centering
	\ifthenelse{\boolean{buildtikzpics}}%
	{%
%		\tikzset{external/export next=false}
		\tikzsetnextfilename{alpha_alphaIniVary_withInset_tRamp3p0_tend100_L12_chi500_dt0p01}
		\begin{tikzpicture}
			\begin{groupplot}%
			[%
				group style =%
				{%
					group size			=	2 by 1,%
					vertical sep		=	0.25em,%
					horizontal sep		=	4.5em,%
					x descriptions at	=	edge bottom,%
				},%
				height		=	0.2\textheight,%
				width		=	0.5\textwidth-3.33pt,%
				xlabel		=	{time $t$},%
				xticklabels	=	{$0$, $ $, $20$, $40$, $60$, $80$, $100$},%
				xtick		=	{0,3,20,40,60,80,100},%
			]%
				\nextgroupplot%
				[%
					ylabel				=	{$\lvert \alpha \rvert\vphantom{\rho^{-5}}$},%
					yticklabels			=	{$\hphantom{-0.}0$, $0.02$, $0.04$, $0.06$, $0.08$},%
					ytick				=	{0, 0.02, 0.04, 0.06, 0.08},%
					scaled y ticks 		=	false,%
					ylabel style 		=	{yshift = -0.2em},%
				]%
					\coordinate (l1) at (rel axis cs:-0.25,0.85);%
					\addplot%
					[%
						color	=	colorA,%
						thick,%
					]%
						table%
						[%
							x expr = \thisrowno{0},%
							y expr = sqrt(\thisrowno{1}*\thisrowno{1}+\thisrowno{2}*\thisrowno{2})%
						]%
							{data/tRamp_3.0/tend_100/tperp_0.2/L_12/U_-4.0/V_-0.25/chi_500/dt_0.01/alphaIni_0.0001/Data_tevol.txt};%
					\addplot%
					[%
						color	=	colorB,%
						densely dashed,
						thick,%
					]%
						table%
						[%
							x expr = \thisrowno{0},%
							y expr = sqrt(\thisrowno{1}*\thisrowno{1}+\thisrowno{2}*\thisrowno{2})%
						]%
							{data/tRamp_3.0/tend_100/tperp_0.2/L_12/U_-4.0/V_-0.25/chi_500/dt_0.01/alphaIni_0.00001/Data_tevol.txt};%
					\addplot%
					[%
						color	=	colorC,%
						densely dotted,
						thick,%
					]%
						table%
						[%
							x expr = \thisrowno{0},%
							y expr = sqrt(\thisrowno{1}*\thisrowno{1}+\thisrowno{2}*\thisrowno{2})%
						]%
							{data/tRamp_3.0/tend_100/tperp_0.2/L_12/U_-4.0/V_-0.25/chi_500/dt_0.01/alphaIni_0.000001/Data_tevol.txt};%
					\coordinate (inset) at (axis description cs:0.15,0.95);%
				\nextgroupplot%
				[%
					ylabel			=	{$\varphi\left( \alpha \right) /\pi\vphantom{\rho^{-5}}$},%
					ylabel style	=	{yshift = -0.2em},%
					yticklabels		=	{$-1$, $-0.8$, $-0.6$, $-0.4$, $-0.2$, $\hphantom{-0.}0$, $1$},%
					ytick			=	{-1, -0.8, -0.6, -0.4, -0.2, 0, 1},%
					legend pos			=	north west,%
					legend cell align	=	{left},%
					legend style =%
					{%
						draw	=	none,%
						fill	=	none,%
						xshift	=	1.25em,
					},%
				]%
					\coordinate (l2) at (rel axis cs:-0.25,0.85);%
					\addplot%
					[%
						color	=	colorA,%
						thick,%
					]%
						table%
						[%
							x expr = \thisrowno{0},%
							y expr = %
							(%
								\thisrowno{1} > 0 ? %
								(%
									rad(atan(\thisrowno{2}/\thisrowno{1}))%
								)%
								:%
								(%
									\thisrowno{2} > 0 ?%
									(%
										(180+atan(\thisrowno{2}/\thisrowno{1}))%
									)%
									:%
									(%
										rad(-180+atan(\thisrowno{2}/\thisrowno{1}))%
									)%
								)%
							)/(pi),%
						]%
							{data/tRamp_3.0/tend_100/tperp_0.2/L_12/U_-4.0/V_-0.25/chi_500/dt_0.01/alphaIni_0.0001/Data_tevol.txt};%
					\addlegendentry{$\alpha_{\mathrm{ini}}=0.01$};
					\addplot%
					[%
						color	=	colorB,%
						densely dashed,%
						thick,%
					]%
						table%
						[%
							x expr = \thisrowno{0},%
							y expr = %
							(%
								\thisrowno{1} > 0 ? %
								(%
									rad(atan(\thisrowno{2}/\thisrowno{1}))%
								)%
								:%
								(%
									\thisrowno{2} > 0 ?%
									(%
										(180+atan(\thisrowno{2}/\thisrowno{1}))%
									)%
									:%
									(%
										rad(-180+atan(\thisrowno{2}/\thisrowno{1}))%
									)%
								)%
							)/(pi),%
						]%
							{data/tRamp_3.0/tend_100/tperp_0.2/L_12/U_-4.0/V_-0.25/chi_500/dt_0.01/alphaIni_0.00001/Data_tevol.txt};%
					\addlegendentry{$\alpha_{\mathrm{ini}}=0.001$};
					\addplot%
					[%
						color	=	colorC,%
						densely dotted,%
						thick,%
					]%
						table%
						[%
							x expr = \thisrowno{0},%
							y expr = %
							(%
								\thisrowno{1} > 0 ? %
								(%
									rad(atan(\thisrowno{2}/\thisrowno{1}))%
								)%
								:%
								(%
									\thisrowno{2} > 0 ?%
									(%
										rad(180+atan(\thisrowno{2}/\thisrowno{1}))%
									)%
									:%
									(%
										rad(-180+atan(\thisrowno{2}/\thisrowno{1}))%
									)%
								)%
							)/(pi),%
						]%
							{data/tRamp_3.0/tend_100/tperp_0.2/L_12/U_-4.0/V_-0.25/chi_500/dt_0.01/alphaIni_0.000001/Data_tevol.txt};%
					\addlegendentry{$\alpha_{\mathrm{ini}}=0.0001$};
			\end{groupplot}%
			\begin{axis}%
			[%
				at				=	(inset),%
				anchor			=	north west,%
				xlabel			=	{\footnotesize $\alpha_{\mathrm{ini}}$},%
				ylabel			=	{\footnotesize $t_{\mathrm{SC}}$},%
				width			=	0.185\textwidth+0.1em,%
				height			=	0.125\textheight,%
				xmin			=	1e-5,
				xmax			=	1e-1,
				ymin			=	30,
				ymax			=	75,
				yticklabels		=	{\footnotesize $50$, \footnotesize $75$},%
				ytick			=	{50, 75},%
				ylabel style	=	{yshift = -0.4em},%
				xticklabels		=	{\footnotesize $10^{-4}$, \footnotesize $10^{-2}$},%
				xtick			=	{1e-4, 1e-2},%
				xlabel style	=	{yshift = 0.2em},%
				scaled y ticks 	=	false,%
				xmode			=	log,
			]%
				\addplot%
				[%
					color	=	black,%
					mark	=	o,
					only marks,
					thick,%
				]%
					table%
					[%
						x expr = \thisrowno{0}*100,%
						y expr = \thisrowno{1}%
					]%
						{data/tRamp_3.0/tend_100/tperp_0.2/localMaxima.dat};%
				\addplot%
				[%
					color	=	gray,%
					domain	=	1e-5:1e-1,
					thick,%
				]%
					{logFct(-4.63609, 0, 0.0143519)};
			\end{axis}%
			\node at (l1) {\subfloat[\label{fig:AbsAlpha_tevol_L12_chi500_dt0.01_alphaIniVary}]{}};%
			\node at (l2) {\subfloat[\label{fig:PhaseAlpha_tevol_L12_chi500_dt0.01_alphaIniVary}]{}};%
		\end{tikzpicture}
	}%
	{%
		\includegraphics{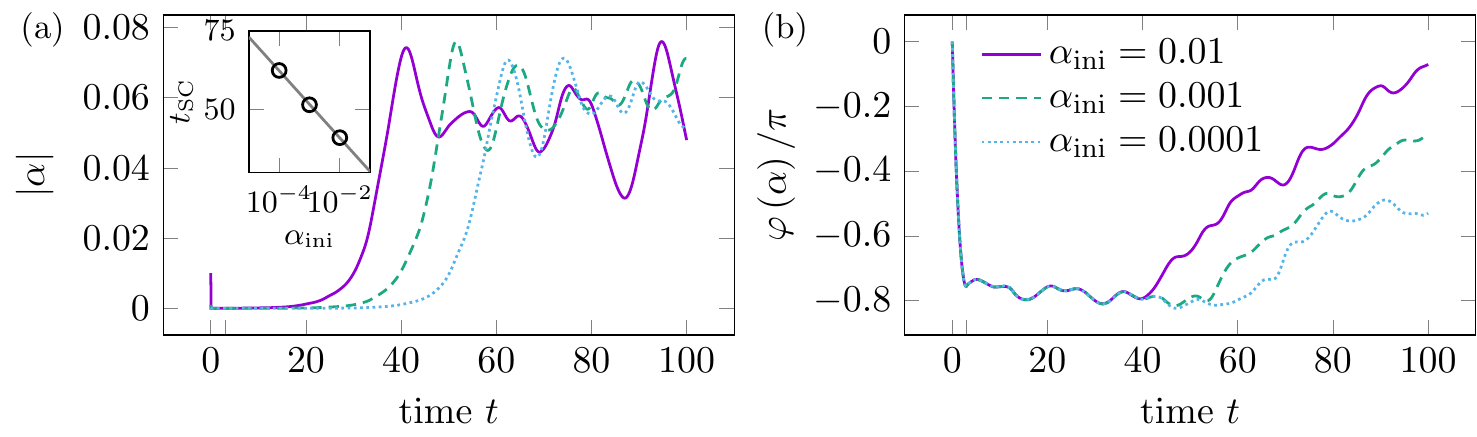}
		
		{\phantomsubcaption\label{fig:AbsAlpha_tevol_L12_chi500_dt0.01_alphaIniVary}}%
		{\phantomsubcaption\label{fig:PhaseAlpha_tevol_L12_chi500_dt0.01_alphaIniVary}}%
	}%
    \caption
    {
    	\label{fig:alpha_tevol_L12_chi500_dt0.01_alphaIniVary}
    	Evolution of the \gls{MF} parameter $\alpha$ split up into its magnitude \protect\subfigref{fig:AbsAlpha_tevol_L12_chi500_dt0.01_alphaIniVary} and its phase \protect\subfigref{fig:PhaseAlpha_tevol_L12_chi500_dt0.01_alphaIniVary} for different initial guesses $\alpha_\mathrm{ini}$ in a $12$\hyp site system. 
    	We see that the reduction of $\alpha_\mathrm{ini}$ induces merely a shift in the data, at least up to time $t_\mathrm{SC}$ at which the first maximum of $\left| \alpha\right|$ occurs. 
    	The inset in \protect\subfigref{fig:AbsAlpha_tevol_L12_chi500_dt0.01_alphaIniVary} shows $t_\mathrm{SC}$ vs. $\alpha_\mathrm{ini}$ and a linear fit on a semilogarithmic scale. 
    	This shows that $t_\mathrm{SC}$ grows merely logarithmically with $\alpha_\mathrm{ini}$. 
    	The data shown was obtained with ${\chi=500}$ and ${\mathrm{d}t = 0.01}$, a ramp time window $\Delta t_{\mathrm{ramp}}=3$, and $\mu$ was taken from \cref{tab:chemical_potential_U-4.0}.
	}
\end{figure}

In the following, we investigate the time evolution of the \gls{BCS} order parameter $\alpha(t)$ (see \cref{eq:self-cons-param-pairing_simplified}) and of the total energy $E(t)$ of the system. 
The latter cannot be expected to remain constant as the \gls{MF} term changes the Hamiltonian \cref{eq:final_eff_MF_Ham_simplified} during evolution.
In addition, we monitor the total density of the system, which should stay at a value of ${\rho = 1}$ (half filling) during the whole time evolution. 
Since we find fast ramps to have lower errors over the simulated time windows than instantaneous quenches, we linearly decrease the value of the nearest\hyp neighbor interaction $V$ from ${V = 0.25}$ to ${V = -0.25}$ within a time window of ${\Delta t_\mathrm{ramp} = 3.0}$. 
A more detailed discussion of the effect of the size of the time window $\Delta t_\mathrm{ramp}$ can be found in \cref{sec:appendix_rampTime}.
\begin{figure}[!t]
    \centering
	\ifthenelse{\boolean{buildtikzpics}}%
	{%
%		\tikzset{external/export next=false}
		\tikzsetnextfilename{tRamp3p0_tend50_L30_chi250_alphaIni0p0001}
		\begin{tikzpicture}
			\begin{groupplot}%
			[%
				group style =%
				{%
					group size			=	2 by 3,%
					vertical sep		=	0.25em,%
					horizontal sep		=	4.5em,%
					x descriptions at	=	edge bottom,%
				},%
				height		=	0.2\textheight,%
				width		=	0.5\textwidth-3.33pt,%
				xlabel		=	{time $t$},%
				xticklabels	=	{$0$, $3$, $20$, $40$},%
				xtick		=	{0,3,20,40},%
			]%
				\nextgroupplot%
				[%
					ylabel				=	{$\lvert \alpha \rvert\vphantom{\rho^{-5}}$},%
					legend pos			=	north west,%
					legend cell align	=	{left},%
					legend style =%
					{%
						draw	=	none,%
						fill	=	none,%
					},%
					yticklabels			=	{$\hphantom{-0.}0$,$0.02$,$0.04$,$0.06$},%
					ytick				=	{0,0.02,0.04,0.06},%
					scaled y ticks 		=	false,%
					ylabel style 		=	{yshift = -0.2em},%
				]%
					\coordinate (l1) at (rel axis cs:-0.25,0.85);%
					\addplot%
					[%
						color	=	colorA,%
						thick,%
					]%
						table%
						[%
							x expr = \thisrowno{0},%
							y expr = sqrt(\thisrowno{1}*\thisrowno{1}+\thisrowno{2}*\thisrowno{2})%
						]%
							{data/tRamp_3.0/tend_50/tperp_0.2/L_30/U_-4.0/V_-0.25/chi_250/dt_0.005/alphaIni_0.0001/Data_tevol.txt};%
					\addlegendentry{$dt=0.005$};%
					\addplot%
					[%
						color	=	colorB,%
						densely dashed,%
						thick,%
					]%
						table%
						[%
							x expr = \thisrowno{0},%
							y expr = sqrt(\thisrowno{1}*\thisrowno{1}+\thisrowno{2}*\thisrowno{2})%
						]%
							{data/tRamp_3.0/tend_50/tperp_0.2/L_30/U_-4.0/V_-0.25/chi_250/dt_0.01/alphaIni_0.0001/Data_tevol.txt};%
					\addlegendentry{$dt=0.01$};%
					\addplot%
					[%
						color	=	colorC,%
						densely dotted,%
						thick,%
					]%
						table%
						[%
							x expr = \thisrowno{0},%
							y expr = sqrt(\thisrowno{1}*\thisrowno{1}+\thisrowno{2}*\thisrowno{2})%
						]%
							{data/tRamp_3.0/tend_50/tperp_0.2/L_30/U_-4.0/V_0.25/chi_250/dt_0.01/alphaIni_0.0001/Data_tevol.txt};%
					\addlegendentry{no quench};%
				\nextgroupplot%
				[%
					ylabel			=	{$E/L\vphantom{\rho^{-5}}$},%
					ylabel style	=	{yshift = -0.2em},%
				]%
					\coordinate (l2) at (rel axis cs:-0.25,0.85);%
					\addplot%
					[%
						color	=	colorA,%
						thick,%
					]%
						table%
						[%
							x expr = \thisrowno{0},%
							y expr = \thisrowno{3}/30.0%
						]%
							{data/tRamp_3.0/tend_50/tperp_0.2/L_30/U_-4.0/V_-0.25/chi_250/dt_0.005/alphaIni_0.0001/Data_tevol.txt};%
					\addplot%
					[%
						color	=	colorB,%
						densely dashed,%
						thick,%
					]%
						table%
						[%
							x expr = \thisrowno{0},%
							y expr = \thisrowno{3}/30.0%
						]%
							{data/tRamp_3.0/tend_50/tperp_0.2/L_30/U_-4.0/V_-0.25/chi_250/dt_0.01/alphaIni_0.0001/Data_tevol.txt};%
					\addplot%
					[%
						color	=	colorC,%
						densely dotted,%
						thick,%
					]%
						table%
						[%
							x expr = \thisrowno{0},%
							y expr = \thisrowno{3}/30.0%
						]%
							{data/tRamp_3.0/tend_50/tperp_0.2/L_30/U_-4.0/V_0.25/chi_250/dt_0.005/alphaIni_0.0001/Data_tevol.txt};%
					\coordinate (inset) at (axis description cs:0.95,0.89);%
				\nextgroupplot%
				[%
					ylabel			=	{$\varphi\left( \alpha \right) /\pi\vphantom{\rho^{-5}}$},%
					ylabel style	=	{yshift = -0.2em},%
					yticklabels		=	{$-1$, $\hphantom{-0.}0$, $1$},%
					ytick			=	{-1, 0, 1},%
				]%
					\coordinate (l3) at (rel axis cs:-0.25,0.85);%
					\addplot%
					[%
						color	=	colorA,%
						thick,%
					]%
						table%
						[%
							x expr = \thisrowno{0},%
							y expr = %
							(%
								\thisrowno{1} > 0 ? %
								(%
									rad(atan(\thisrowno{2}/\thisrowno{1}))%
								)%
								:%
								(%
									\thisrowno{2} > 0 ?%
									(%
										(180+atan(\thisrowno{2}/\thisrowno{1}))%
									)%
									:%
									(%
										rad(-180+atan(\thisrowno{2}/\thisrowno{1}))%
									)%
								)%
							)/(pi),%
						]%
							{data/tRamp_3.0/tend_50/tperp_0.2/L_30/U_-4.0/V_-0.25/chi_250/dt_0.005/alphaIni_0.0001/Data_tevol.txt};%
					\addplot%
					[%
						color	=	colorB,%
						densely dashed,%
						thick,%
					]%
						table%
						[%
							x expr = \thisrowno{0},%
							y expr = %
							(%
								\thisrowno{1} > 0 ? %
								(%
									rad(atan(\thisrowno{2}/\thisrowno{1}))%
								)%
								:%
								(%
									\thisrowno{2} > 0 ?%
									(%
										(180+atan(\thisrowno{2}/\thisrowno{1}))%
									)%
									:%
									(%
										rad(-180+atan(\thisrowno{2}/\thisrowno{1}))%
									)%
								)%
							)/(pi),%
						]%
							{data/tRamp_3.0/tend_50/tperp_0.2/L_30/U_-4.0/V_-0.25/chi_250/dt_0.01/alphaIni_0.0001/Data_tevol.txt};%
					\addplot%
					[%
						color	=	colorC,%
						densely dotted,%
						thick,%
					]%
						table%
						[%
							x expr = \thisrowno{0},%
							y expr = %
							(%
								\thisrowno{1} > 0 ? %
								(%
									rad(atan(\thisrowno{2}/\thisrowno{1}))%
								)%
								:%
								(%
									\thisrowno{2} > 0 ?%
									(%
										rad(180+atan(\thisrowno{2}/\thisrowno{1}))%
									)%
									:%
									(%
										rad(-180+atan(\thisrowno{2}/\thisrowno{1}))%
									)%
								)%
							)/(pi),%
						]%
							{data/tRamp_3.0/tend_50/tperp_0.2/L_30/U_-4.0/V_0.25/chi_250/dt_0.005/alphaIni_0.0001/Data_tevol.txt};%
				\nextgroupplot%
				[%
					ylabel			=	{$\left( 1-\rho \right)/10^{-5}$},%
					yticklabels		=	{$\hphantom{-0.}0$, $1$, $2$, $3$},%
					ytick			=	{0, 1, 2, 3},%
					ylabel style	=	{yshift = -0.2em},%
				]%
					\coordinate (l4) at (rel axis cs:-0.25,0.85);%
					\addplot%
					[%
						color	=	colorA,%
						thick,%
					]%
						table%
						[%
							x expr = \thisrowno{0},%
							y expr = \eval{10^5-\thisrowno{15}*10^5}%
						]%
							{data/tRamp_3.0/tend_50/tperp_0.2/L_30/U_-4.0/V_-0.25/chi_250/dt_0.005/alphaIni_0.0001/Data_tevol.txt};%
					\addplot%
					[%
						color	=	colorB,%
						densely dashed,%
						thick,%
					]%
						table%
						[%
							x expr = \thisrowno{0},%
							y expr = \eval{10^5-\thisrowno{15}*10^5}%
						]%
							{data/tRamp_3.0/tend_50/tperp_0.2/L_30/U_-4.0/V_-0.25/chi_250/dt_0.01/alphaIni_0.0001/Data_tevol.txt};%
					\addplot%
					[%
						color	=	colorC,%
						densely dotted,%
						thick,%
					]%
						table%
						[%
							x expr = \thisrowno{0},%
							y expr = \eval{10^5-\thisrowno{15}*10^5}%
						]%
							{data/tRamp_3.0/tend_50/tperp_0.2/L_30/U_-4.0/V_0.25/chi_250/dt_0.005/alphaIni_0.0001/Data_tevol.txt};%
				\nextgroupplot%
				[%
					ylabel			=	{$V\vphantom{\rho^{-5}}$},%
					height			=	0.1125\textheight,%
					ylabel style	=	{yshift = -0.2em},%
				]%
					\coordinate (l5) at (rel axis cs:-0.25,0.55);%
					\addplot%
					[%
						color	=	black,%
						thick,%
					]%
						table%
						[%
							x expr = \thisrowno{0},%
							y expr = \thisrowno{1}%
						]%
							{data/tRamp_3.0/tend_50/tperp_0.2/L_30/U_-4.0/V_-0.25/chi_250/dt_0.005/alphaIni_0.0001/V_in_time.txt};%
				\nextgroupplot%
				[%
					ylabel			=	{$V\vphantom{\rho^{-5}}$},%
					height			=	0.1125\textheight,%
					ylabel style	=	{yshift = -0.2em},%
				]%
					\coordinate (l6) at (rel axis cs:-0.25,0.55);%
					\addplot%
					[%
						color	=	black,%
						thick,%
					]%
						table%
						[%
							x expr = \thisrowno{0},%
							y expr = \thisrowno{1}%
						]%
							{data/tRamp_3.0/tend_50/tperp_0.2/L_30/U_-4.0/V_-0.25/chi_250/dt_0.005/alphaIni_0.0001/V_in_time.txt};%
			\end{groupplot}%
			\begin{axis}%
			[%
				at			=	(inset),%
				anchor		=	north east,%
				xmin		=	3,%
				xlabel		=	{\footnotesize time $t$},%
				ylabel		=	{\footnotesize E/L},%
				width		=	0.275\textwidth,%
				height		=	0.125\textheight,%
				yticklabels	=	{\footnotesize $-0.225$, \footnotesize $-0.22$, \footnotesize $-0.215$},%
				ytick		=	{-0.225, -0.22, -0.215},%
				xticklabels	=	{\footnotesize $3$, \footnotesize $20$, \footnotesize $40$},%
				xtick		=	{3,20,40},%
			]%
				\addplot%
				[%
					color	=	colorA,%
					thick,%
				]%
					table%
					[%
						x expr = \thisrowno{0},%
						y expr = \thisrowno{3}/30.0%
					]%
						{data/tRamp_3.0/tend_50/tperp_0.2/L_30/U_-4.0/V_-0.25/chi_250/dt_0.005/alphaIni_0.0001/Data_tevol.txt};%
				\addplot%
				[%
					color	=	colorB,%
					densely dashed,%
					thick,%
				]%
					table%
					[%
						x expr = \thisrowno{0},%
						y expr = \thisrowno{3}/30.0%
					]%
						{data/tRamp_3.0/tend_50/tperp_0.2/L_30/U_-4.0/V_-0.25/chi_250/dt_0.01/alphaIni_0.0001/Data_tevol.txt};%
			\end{axis}%
			\node at (l1) {\subfloat[\label{fig:AbsAlpha_tevol_L30_chi250}]{}};%
			\node at (l3) {\subfloat[\label{fig:PhaseAlpha_tevol_L30_chi250}]{}};%
			\node at (l5) {\subfloat[\label{fig:V_alpha_tevol_L30_chi250}]{}};%
			\node at (l2) {\subfloat[\label{fig:E_tevol_L30_chi250}]{}};%
			\node at (l4) {\subfloat[\label{fig:Rho_tevol_L30_chi250}]{}};%
			\node at (l6) {\subfloat[\label{fig:V_E_tevol_L30_chi250}]{}};%
		\end{tikzpicture}
	}%
	{%
		\includegraphics{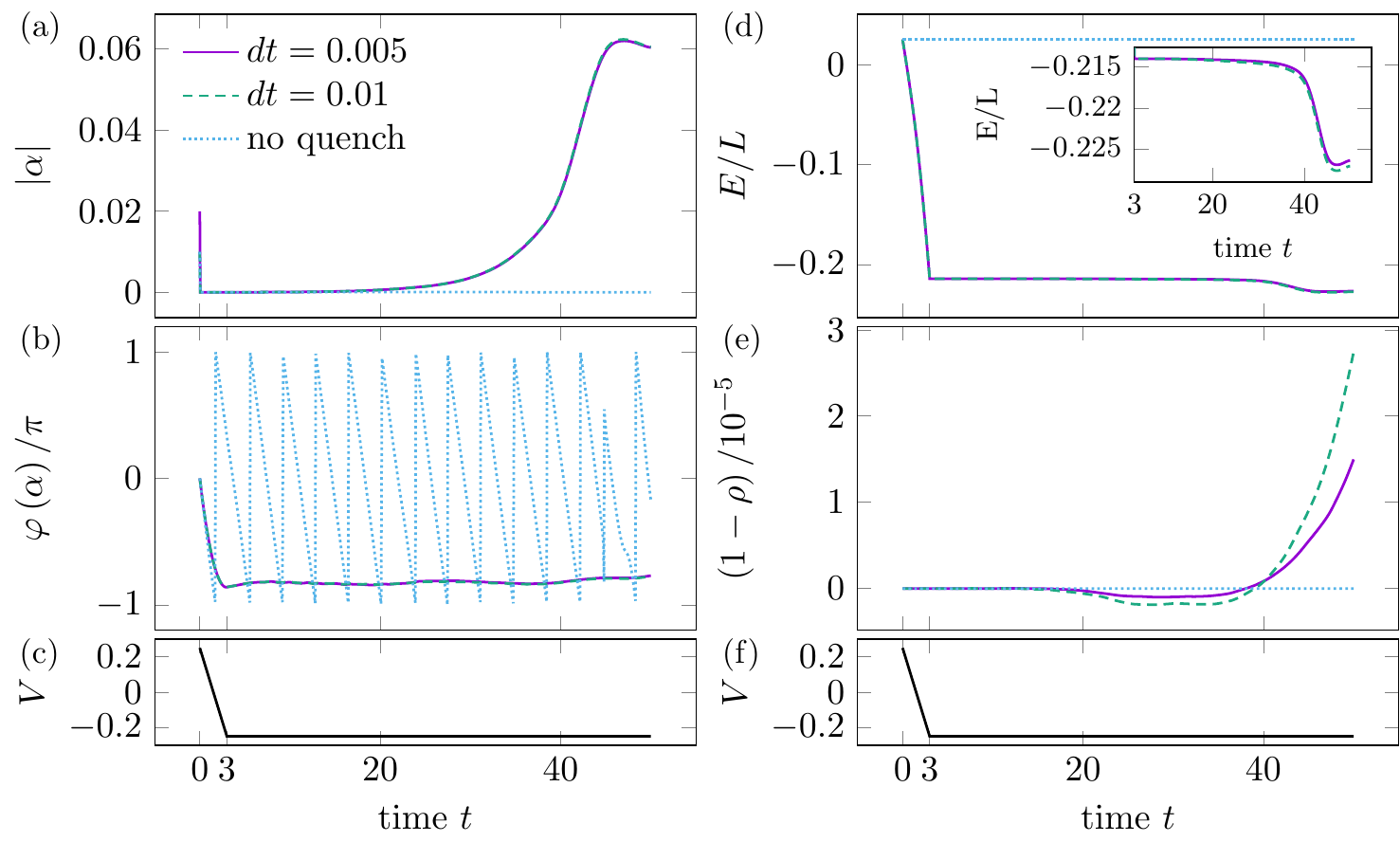}
		
		{\phantomsubcaption\label{fig:AbsAlpha_tevol_L30_chi250}}%
		{\phantomsubcaption\label{fig:PhaseAlpha_tevol_L30_chi250}}%
		{\phantomsubcaption\label{fig:V_alpha_tevol_L30_chi250}}%
		{\phantomsubcaption\label{fig:E_tevol_L30_chi250}}%
		{\phantomsubcaption\label{fig:Rho_tevol_L30_chi250}}%
		{\phantomsubcaption\label{fig:V_E_tevol_L30_chi250}}%
	}%
    \caption
    {
	    \label{fig:alphaEandRho_tevol_L30_chi250}
    	Evolution of the considered parameters in time during and after a ramp on a $30$\hyp site system. 
    	The plots at the bottom (\protect\subfigref{fig:V_alpha_tevol_L30_chi250} and \protect\subfigref{fig:V_E_tevol_L30_chi250}) show the nearest\hyp neighbor interaction, which decreases from ${V = 0.25}$ to ${V = -0.25}$ during a time window of ${\Delta t_\mathrm{ramp} = 3.0}$.
		Evolution of the \gls{MF} parameter $\alpha$ split up into magnitude \protect\subfigref{fig:AbsAlpha_tevol_L30_chi250} and phase \protect\subfigref{fig:PhaseAlpha_tevol_L30_chi250}.
		Evolution of the total energy per site of the system \protect\subfigref{fig:E_tevol_L30_chi250} and the total density \protect\subfigref{fig:Rho_tevol_L30_chi250}.
		The inset in \protect\subfigref{fig:E_tevol_L30_chi250} shows the evolution of the energy per site after $V$ was decreased. 
		The legend is valid for all plots.
		All the data shown here were obtained with a bond dimension of $\chi = 250$, an initial guess of the \gls{MF} parameter of $\alpha_\mathrm{ini} = 10^{-4}/\mathrm{d}t$, and the chemical potential was taken from \cref{tab:chemical_potential_U-4.0}. 
		We compare the ramp scenario (solid violet and dashed green) with an evolution during which we keep the nearest neighbor interaction at $V = 0.25$ constant (dotted blue). 
		For the latter calculation we chose a time step of $\mathrm{d}t = 0.01$.
	}
\end{figure}
In \cref{fig:alphaEandRho_tevol_L30_chi250} we see the results for a $30$\hyp site system for an evolution up to time ${t_\mathrm{end} = 50}$. 
Since $\alpha(t)$ is complex valued we show the evolution of the magnitude $|\alpha(t)|$ and of the phase $\varphi(t)$ of the order parameter in \cref{fig:alphaEandRho_tevol_L30_chi250,fig:alphaEandRho_tevol_Lvary_chi500,fig:alpha_tevol_L12_chi500_dt0.01_alphaIniVary}.
We find that $|\alpha(t)|$ grows up to time ${t \sim 45}$ to a value of approximately ${\left|\alpha\right| \approx 0.06}$, which is clearly non vanishing and hence indicates the formation of a non\hyp equilibrium \gls{SC} state. 
In contrast, if we consider a time evolution without a quench or ramp, i.e., ${V = 0.25}$ during the whole evolution, the value of $\alpha$ stays unchanged at an order of magnitude of $10^{-5}$ throughout the whole time evolution as can be seen by the dotted blue lines in \cref{fig:alphaEandRho_tevol_L30_chi250}. 
The phase $\varphi(t)$ decreases as long as $V$ is decreasing, then oscillates around a value of approximately ${\varphi(\alpha)/\pi \approx -0.8}$ and seems to increase again slightly when $\left|\alpha\right|$ has reached its maximum. 
We interpret this behavior as an expression of a Josephson effect in\hyp between \gls{1D} chains to the extent it can be captured by a single \gls{1D} system with time-evolving \gls{MF} amplitudes. 
As a kernel of \gls{SC} order manifests itself in the different chains of the \gls{2D} array the macroscopic phases of \gls{SC} states, within each chain, will be initially uncorrelated, then start aligning via the Josephson effect.
With density fluctuating within each individual chain the Josephson effect will keep the phase fluctuating while the system finds a new equilibrium after the rapid ramp, as \cref{fig:PhaseAlpha_tevol_L30_chi250} shows.
In \cref{fig:E_tevol_L30_chi250} we show the evolution of the total energy per site $E(t)/L$ and in \cref{fig:Rho_tevol_L30_chi250} the deviation of the total density $\rho(t)$ from the desired value ${\rho_\mathrm{target} = 1}$.
We find that this deviation is of the order of ${3 \cdot 10^{-5}}$ or smaller for all the times treated, indicating that keeping the chemical potential $\mu$ fixed leads only to small errors. 
The total energy per site $E/L$ behaves as expected during the ramp and decreases almost linearly for the duration of the ramp. 
Afterwards, we first observe a nearly constant behavior, then a strong decrease until a minimum at time ${t \approx 45}$, shown in the inset of \cref{fig:AbsAlpha_tevol_L30_chi250}. 
We read the behavior of $E(t)/L$, especially at long times, as the system starting to further lower its energy through condensing Cooper pairs, as the drop in $E(t)/L$ coincides markedly with the onset of a finite value of $\alpha(t)$.
We also study the effect of system size, to make certain the dynamical onset of superconductivity would survive in the thermodynamic limit. 
In \cref{fig:alphaEandRho_tevol_Lvary_chi500} we compare the results for different chain lengths $L$. 
We obtain a shift of the instant $t_\mathrm{SC}$, at which $|\alpha(t)|$ reaches its first maximum. 
The data of the $12$\hyp site system shows the onset of oscillation for $\left|\alpha(t)\right|$ around a finite value, indicating a dynamically induced \gls{SC} phase (longer\hyp time simulations for $L=12$ further confirm this, as shown in \cref{fig:alpha_tevol_L12_chi500_dt0.01_alphaIniVary,fig:error_L12,fig:error_L30} for times up to ${t_\mathrm{max}=100}$). 
The inset of \cref{fig:AbsAlpha_tevol_L12_chi500_dt0.01_alphaIniVary} displays an extrapolation in inverse chain length $1/L$ of $t_\mathrm{SC}$.
In order to see whether $t_{\mathrm{SC}}$ diverges we performed a quadratic and a linear fit, both indicating a finite value in the limit $L\rightarrow\infty$.
Since for the larger system sizes $|\alpha(t)|$ starts to oscillate at around the maximal time reached by us, it is difficult to obtain a finite\hyp size extrapolation of the value of the \gls{SC} order parameter.
In order to do so, one needs to extend the simulations for the larger systems to substantially longer times, which is beyond the scope of this paper. 
\subsection{\label{sec:accuracyOfData}Accuracy and sensitivity of the results on the simulation parameters}
The results so far were all obtained using the same parameters for the self\hyp consistency cycle.
The question arises, how sensitive the results depend on parameters like the initial guess of the \gls{MF} parameter $\alpha_\mathrm{ini}$ (see \cref{sec:self-cons-tevol_algorithm}), the bond dimension of the \gls{MPS} calculations, or the discrete time step $\mathrm{d}t$. 
To study these effects, we focus on the $12$\hyp site system in order to reach the longest time scales.
\Cref{fig:alpha_tevol_L12_chi500_dt0.01_alphaIniVary} shows the evolution of the magnitude and phase of $\alpha(t)$ for different initial values $\alpha_\mathrm{ini}$. 
Decreasing the value of $\alpha_\mathrm{ini}$ induces a shift of $t_\mathrm{SC}$ to later times. 
In order to further analyze this, we plot the value of $t_\mathrm{SC}$ against the value of $\alpha_\mathrm{ini}$ in the inset of \cref{fig:AbsAlpha_tevol_L12_chi500_dt0.01_alphaIniVary}. 
Speaking to the soundness of our \gls{MF} approximation, we find that $t_{SC}$ increases only very weakly with $\alpha_\mathrm{ini}$, i.e., logarithmically. 
While this indicates a diverging time for the onset of \gls{SC} order in the limit ${\alpha_\mathrm{ini} \rightarrow 0}$, this is merely consistent with ${\alpha_\mathrm{ini} = 0}$ being an unstable fix point of the dynamic \gls{MF} algorithm in the regime we ramp into. 
But any finite value, even a microscopic one, will yield dynamically induced \gls{SC} order in finite time when ramping into the \gls{SC} parameter regime. 
As argued at the outset of \cref{sec:resultsAndDiscussion}: on general physical grounds there will always be some electron pairs whose center\hyp of\hyp mass momentum is zero.
In \cref{fig:alphaEandRho_tevol_L30_chi250} we compare two different discretized time steps, ${\mathrm{d}t = 0.005}$ and ${\mathrm{d}t=0.01}$, respectively.
The results are nearly identical, only a small deviation of the total density, which agrees up to ${\sim 10^{-5}}$, can be seen in \cref{fig:Rho_tevol_L30_chi250}.

Next, we check the accuracy of our results if the \gls{MPS} bond dimension $\chi$ is changed.
This additional check is necessary since the discarded weight is already below $10^{-6}$ for the smallest bond dimension.
For this purpose we compute the deviation of the value of an observable $\mathcal{O}$ for two different values of $\chi$, 
\begin{align}
    \delta_{\mathcal{O},\chi_1,\chi_2} = \left|\mathcal{O}(\chi_1) - \mathcal{O}(\chi_2)\right| \;.
    \label{eq:def_diffBondDim_vary}
\end{align}
At any fixed value of $\alpha_\mathrm{ini}$ and $\mathrm{d}t$ we find this to be the most reliable estimator for the accuracy of our combined \gls{MPS}+\gls{MF} approach (assuming the latter parameter is chosen to be sufficiently small) and focus in the following on this quantity.
\begin{figure}[!t]
    \centering

	\ifthenelse{\boolean{buildtikzpics}}%
	{%
%		\immediate\write18{paste data/tRamp_3.0/tend_100/tperp_0.2/L_12/U_-4.0/V_-0.25/chi_500/dt_0.01/alphaIni_0.0001/Data_tevol.txt data/tRamp_3.0/tend_100/tperp_0.2/L_12/U_-4.0/V_-0.25/chi_1000/dt_0.01/alphaIni_0.0001/Data_tevol.txt > data/tRamp_3.0/tend_100/tperp_0.2/L_12/U_-4.0/V_-0.25/combined_Data_tevol.txt}
%		\tikzset{external/export next=false}
		\tikzsetnextfilename{diffAlpha_diffE_tRamp3p0_tend100_L12_dt0p01_alphaIni0p0001}
		\begin{tikzpicture}
			\begin{groupplot}%
			[%
				group style =%
				{%
					group size			=	2 by 2,%
					vertical sep		=	0.25em,%
					horizontal sep		=	5em,%
					x descriptions at	=	edge bottom,%
				},%
				height		=	0.2\textheight,%
				width		=	0.5\textwidth-16.5pt,%
				xlabel		=	{time $t$},%
				xmin		=	0,%
				xmax		=	100,%
				enlarge x limits,
				xticklabels	=	{$0$, , $20$, $40$, $60$, $80$, $100$},%
				xtick		=	{0,3,20,40,60,80,100},%
			]%
				\nextgroupplot%
				[%
					ylabel				=	{$\lvert \alpha \rvert \vphantom{\rho^{-5}}$},%
					legend pos			=	south east,%
					legend cell align	=	{left},%
					legend style =%
					{%
						draw	=	none,%
						fill	=	none,%
					},%
					scaled y ticks 		=	false,%
					yticklabels			=	{$\hphantom{-0.}0$,$0.02$,$0.04$,$0.06$,$0.08$},%
					ytick				=	{0,0.02,0.04,0.06,0.08},%
					ylabel style 		=	{yshift = -0.2em},%
				]%
					\coordinate (l1) at (rel axis cs:-0.25,0.85);%
					\addplot%
					[%
						color	=	colorA,%
						thick,%
						restrict x to domain	=	0:100,%
					]%
						table%
						[%
							x expr = \thisrowno{0},%
							y expr = sqrt(\thisrowno{1}*\thisrowno{1}+\thisrowno{2}*\thisrowno{2})%
						]%
							{data/tRamp_3.0/tend_100/tperp_0.2/L_12/U_-4.0/V_-0.25/chi_500/dt_0.01/alphaIni_0.0001/Data_tevol.txt};%
					\addlegendentry{$\chi=500$};%
					\addplot%
					[%
						color	=	colorB,%
						thick,%
						restrict x to domain	=	0:100,%
					]%
						table%
						[%
							x expr = \thisrowno{0},%
							y expr = sqrt(\thisrowno{1}*\thisrowno{1}+\thisrowno{2}*\thisrowno{2})%
						]%
							{data/tRamp_3.0/tend_100/tperp_0.2/L_12/U_-4.0/V_-0.25/chi_1000/dt_0.01/alphaIni_0.0001/Data_tevol.txt};%
					\addlegendentry{$\chi=1000$};%				
				\nextgroupplot%
				[%
					ylabel				=	{$E/L \vphantom{\rho^{-5}}$},%
					legend pos			=	north west,%
					legend cell align	=	{left},%
					legend style =%
					{%
						draw	=	none,%
						fill	=	none,%
					},%
					scaled y ticks 		=	false,%
					ylabel style 		=	{yshift = -0.2em},%
				]%
					\coordinate (l2) at (rel axis cs:-0.25,0.85);%
					\addplot%
					[%
						color	=	colorA,%
						thick,%
						restrict x to domain	=	3:100,%
					]%
						table%
						[%
							x expr = \thisrowno{0},%
							y expr = \thisrowno{3}/12.0%
						]%
							{data/tRamp_3.0/tend_100/tperp_0.2/L_12/U_-4.0/V_-0.25/chi_500/dt_0.01/alphaIni_0.0001/Data_tevol.txt};%
					\addplot%
					[%
						color	=	colorB,%
						thick,%
						restrict x to domain	=	3:100,%
					]%
						table%
						[%
							x expr = \thisrowno{0},%
							y expr = \thisrowno{3}/12.0%
						]%
							{data/tRamp_3.0/tend_100/tperp_0.2/L_12/U_-4.0/V_-0.25/chi_1000/dt_0.01/alphaIni_0.0001/Data_tevol.txt};%
				\nextgroupplot%
				[%
					ylabel				=	{$\delta_{\lvert \alpha \rvert, 500, 1000}$},%
					scaled y ticks 		=	false,%
					ylabel style 		=	{yshift = -0.2em},%
					ymode 				=	log,%
				]%
					\coordinate (l3) at (rel axis cs:-0.25,0.85);%
					\addplot%
					[%
						color	=	colorC,%
						thick,%
						restrict x to domain	=	0:100,%
					]%
						table%
						[%
							x expr = \thisrowno{0},%
							y expr = \eval{abs(sqrt(\thisrowno{1}*\thisrowno{1}+\thisrowno{2}*\thisrowno{2}) - sqrt(\thisrowno{18}*\thisrowno{18}+\thisrowno{19}*\thisrowno{19}))}%
						]%
							{data/tRamp_3.0/tend_100/tperp_0.2/L_12/U_-4.0/V_-0.25/combined_Data_tevol.txt};%
				\nextgroupplot%
				[%
					ylabel				=	{$\delta_{E/L, 500, 1000}$},%
					scaled y ticks 		=	false,%
					ylabel style 		=	{yshift = -0.2em},%
					ymode 				=	log,%
				]%
					\coordinate (l4) at (rel axis cs:-0.25,0.85);%
					\addplot%
					[%
						color	=	colorC,%
						thick,%
						restrict x to domain	=	3:100,%
					]%
						table%
						[%
							x expr = \thisrowno{0},%
							y expr = \eval{abs(\thisrowno{3}/12.0 - \thisrowno{20}/12.0)}%
						]%
							{data/tRamp_3.0/tend_100/tperp_0.2/L_12/U_-4.0/V_-0.25/combined_Data_tevol.txt};%
			\end{groupplot}%
			\node at (l1) {\subfloat[\label{fig:absAlpha_L12}]{}};%
			\node at (l3) {\subfloat[\label{fig:errorAlpha_L12}]{}};%
			\node at (l2) {\subfloat[\label{fig:E_L12}]{}};%
			\node at (l4) {\subfloat[\label{fig:errorE_L12}]{}};%
		\end{tikzpicture}
	}%
	{%
		\includegraphics{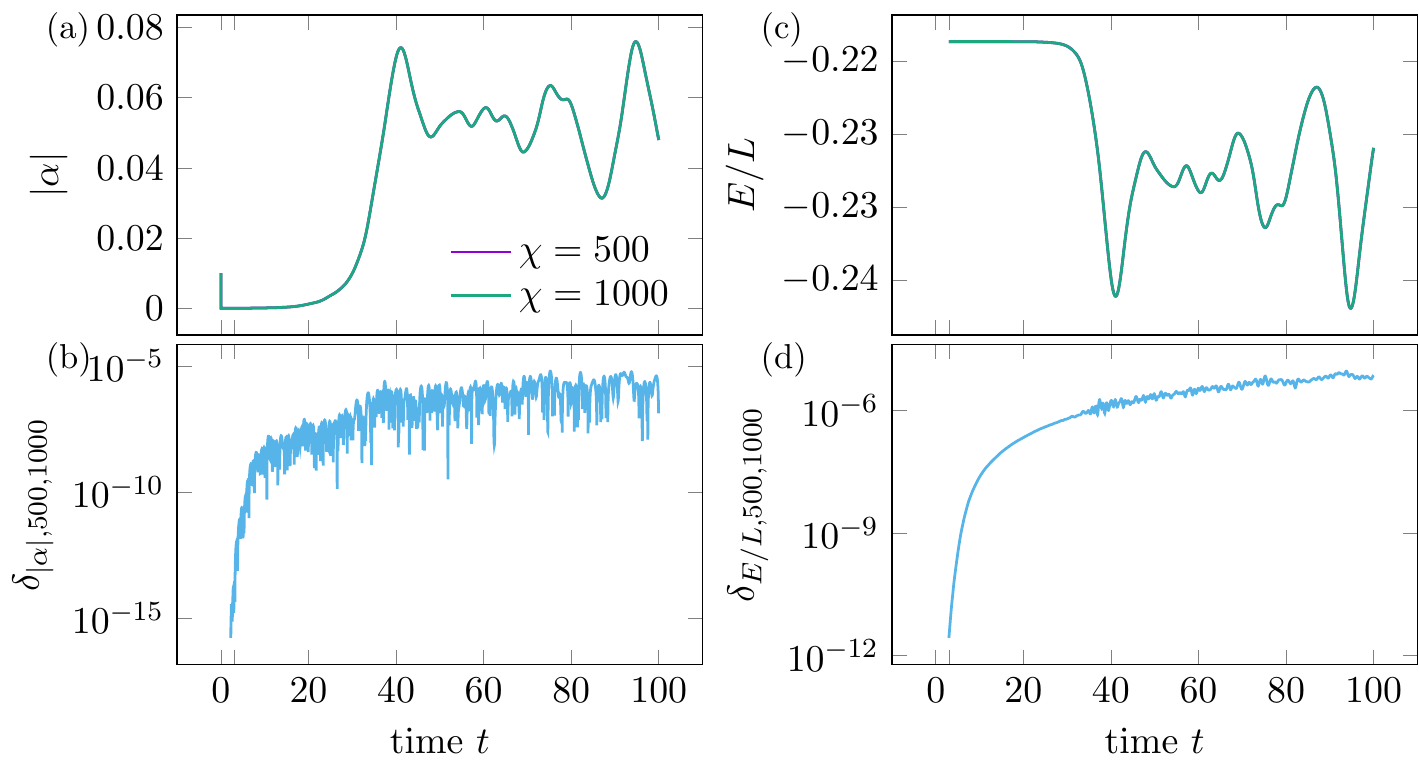}
		
		{\phantomsubcaption\label{fig:absAlpha_L12}}%
		{\phantomsubcaption\label{fig:errorAlpha_L12}}%
		{\phantomsubcaption\label{fig:E_L12}}%
		{\phantomsubcaption\label{fig:errorE_L12}}%
	}%
    \caption
    {
	    \label{fig:error_L12}
    	Evolution of \protect\subfigref{fig:absAlpha_L12} the magnitude of the \gls{MF} parameter $\left|\alpha\right|$ and \protect\subfigref{fig:E_L12} the total energy $E$ for two different bond dimensions ${\chi_1=500}$ and ${\chi_2=1000}$ in a $12$\hyp site system. 
    	\protect\subfigref{fig:errorAlpha_L12} and \protect\subfigref{fig:errorE_L12} show the difference $\delta_{\mathcal{O},\chi_1,\chi_2}$ between the observables we measure for these two different bond dimensions. 
    	All calculations were done with ${\alpha_\mathrm{ini}=10^{-4}/\mathrm{d}t}$, a ramp time window $\Delta t_{\mathrm{ramp}}=3$, and ${\mathrm{d}t=0.01}$.
    }
\end{figure}

\begin{figure}[!t]
	\centering
	\ifthenelse{\boolean{buildtikzpics}}%
	{%
%		\immediate\write18{paste data/tRamp_3.0/tend_50/tperp_0.2/L_30/U_-4.0/V_-0.25/chi_250/dt_0.01/alphaIni_0.0001/Data_tevol.txt data/tRamp_3.0/tend_50/tperp_0.2/L_30/U_-4.0/V_-0.25/chi_500/dt_0.01/alphaIni_0.0001/Data_tevol.txt > data/tRamp_3.0/tend_50/tperp_0.2/L_30/U_-4.0/V_-0.25/combined_Data_tevol.txt}
%		\tikzset{external/export next=false}
		\tikzsetnextfilename{diffAlpha_diffE_tRamp3p0_tend50_L30_dt0p01_alphaIni0p0001}
		\begin{tikzpicture}
			\begin{groupplot}%
			[%
				group style =%
				{%
					group size			=	2 by 2,%
					vertical sep		=	0.25em,%
					horizontal sep		=	5em,%
					x descriptions at	=	edge bottom,%
				},%
				height		=	0.2\textheight,%
				width		=	0.5\textwidth-16.5pt,%
				xlabel		=	{time $t$},%
				xmin		=	0,%
				xmax		=	50,%
				enlarge x limits,
				xticklabels	=	{$0$, , $10$, $20$, $30$, $40$, $50$},%
				xtick		=	{0,3,10,20,30,40,50},%
			]%
				\nextgroupplot%
				[%
					ylabel				=	{$\lvert \alpha \rvert \vphantom{\rho^{-5}}$},%
					legend pos			=	north west,%
					legend cell align	=	{left},%
					legend style =%
					{%
						draw	=	none,%
						fill	=	none,%
					},%
					scaled y ticks 		=	false,%
					yticklabels			=	{$\hphantom{-0.}0$,$0.02$,$0.04$,$0.06$},%
					ytick				=	{0,0.02,0.04,0.06},%
					ylabel style 		=	{yshift = -0.2em},%
				]%
					\coordinate (l1) at (rel axis cs:-0.25,0.85);%
					\addplot%
					[%
						color	=	colorA,%
						thick,%
						restrict x to domain	=	0:50,%
					]%
						table%
						[%
							x expr = \thisrowno{0},%
							y expr = sqrt(\thisrowno{1}*\thisrowno{1}+\thisrowno{2}*\thisrowno{2})%
						]%
							{data/tRamp_3.0/tend_50/tperp_0.2/L_30/U_-4.0/V_-0.25/chi_250/dt_0.01/alphaIni_0.0001/Data_tevol.txt};%
					\addlegendentry{$\chi=250$};%
					\addplot%
					[%
						color	=	colorB,%
						thick,%
						restrict x to domain	=	0:50,%
					]%
						table%
						[%
							x expr = \thisrowno{0},%
							y expr = sqrt(\thisrowno{1}*\thisrowno{1}+\thisrowno{2}*\thisrowno{2})%
						]%
							{data/tRamp_3.0/tend_50/tperp_0.2/L_30/U_-4.0/V_-0.25/chi_500/dt_0.01/alphaIni_0.0001/Data_tevol.txt};%
					\addlegendentry{$\chi=500$};%				
				\nextgroupplot%
				[%
					ylabel				=	{$E/L \vphantom{\rho^{-5}}$},%
					legend pos			=	north west,%
					legend cell align	=	{left},%
					legend style =%
					{%
						draw	=	none,%
						fill	=	none,%
					},%
					scaled y ticks 		=	false,%
					ylabel style 		=	{yshift = -0.2em},%
				]%
					\coordinate (l2) at (rel axis cs:-0.25,0.85);%
					\addplot%
					[%
						color	=	colorA,%
						thick,%
						restrict x to domain	=	3:50,%
					]%
						table%
						[%
							x expr = \thisrowno{0},%
							y expr = \thisrowno{3}/30.0%
						]%
							{data/tRamp_3.0/tend_50/tperp_0.2/L_30/U_-4.0/V_-0.25/chi_250/dt_0.01/alphaIni_0.0001/Data_tevol.txt};%
					\addplot%
					[%
						color	=	colorB,%
						thick,%
						restrict x to domain	=	3:50,%
					]%
						table%
						[%
							x expr = \thisrowno{0},%
							y expr = \thisrowno{3}/30.0%
						]%
							{data/tRamp_3.0/tend_50/tperp_0.2/L_30/U_-4.0/V_-0.25/chi_500/dt_0.01/alphaIni_0.0001/Data_tevol.txt};%
				\nextgroupplot%
				[%
					ylabel				=	{$\delta_{\lvert \alpha \rvert, 250, 500}$},%
					scaled y ticks 		=	false,%
					ylabel style 		=	{yshift = -0.2em},%
					ymode 				=	log,%
				]%
					\coordinate (l3) at (rel axis cs:-0.25,0.85);%
					\addplot%
					[%
						color	=	colorC,%
						thick,%
						restrict x to domain	=	0:50,%
					]%
						table%
						[%
							x expr = \thisrowno{0},%
							y expr = \eval{abs(sqrt(\thisrowno{1}*\thisrowno{1}+\thisrowno{2}*\thisrowno{2}) - sqrt(\thisrowno{18}*\thisrowno{18}+\thisrowno{19}*\thisrowno{19}))}%
						]%
							{data/tRamp_3.0/tend_50/tperp_0.2/L_30/U_-4.0/V_-0.25/combined_Data_tevol.txt};%
				\nextgroupplot%
				[%
					ylabel				=	{$\delta_{E/L, 250, 500}$},%
					scaled y ticks 		=	false,%
					ylabel style 		=	{yshift = -0.2em},%
					ymode 				=	log,%
				]%
					\coordinate (l4) at (rel axis cs:-0.25,0.85);%
					\addplot%
					[%
						color	=	colorC,%
						thick,%
						restrict x to domain	=	3:50,%
					]%
						table%
						[%
							x expr = \thisrowno{0},%
							y expr = \eval{abs(\thisrowno{3}/30.0 - \thisrowno{20}/30.0)}%
						]%
							{data/tRamp_3.0/tend_50/tperp_0.2/L_30/U_-4.0/V_-0.25/combined_Data_tevol.txt};%
			\end{groupplot}%
			\node at (l1) {\subfloat[\label{fig:absAlpha_L30}]{}};%
			\node at (l3) {\subfloat[\label{fig:errorAlpha_L30}]{}};%
			\node at (l2) {\subfloat[\label{fig:E_L30}]{}};%
			\node at (l4) {\subfloat[\label{fig:errorE_L30}]{}};%
		\end{tikzpicture}
	}%
	{%
		\includegraphics{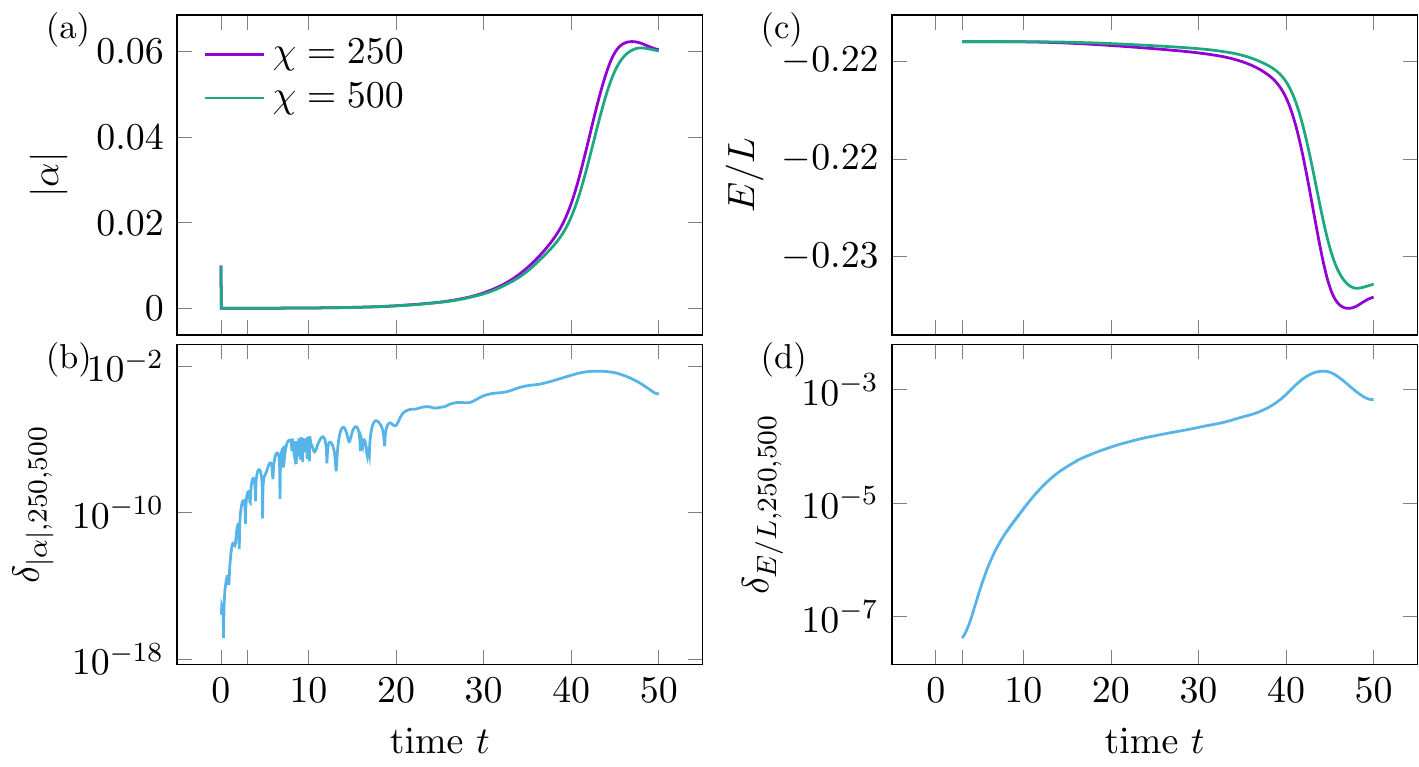}
		
		{\phantomsubcaption\label{fig:absAlpha_L30}}%
		{\phantomsubcaption\label{fig:errorAlpha_L30}}%
		{\phantomsubcaption\label{fig:E_L30}}%
		{\phantomsubcaption\label{fig:errorE_L30}}%
	}%
    \caption
    {
    	\label{fig:error_L30}
    	Analog to \cref{fig:error_L12} but for a $30$\hyp site system and for bond dimensions ${\chi_1=250}$ and ${\chi_2 = 500}$.
    }
\end{figure}
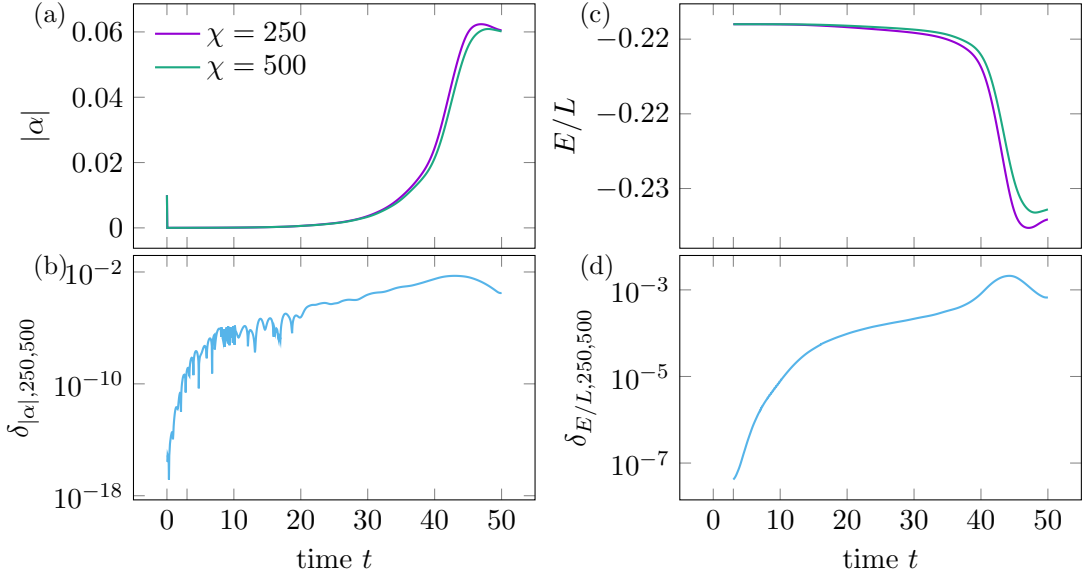
In \cref{fig:error_L12,fig:error_L30} we present results for the observables $\left|\alpha(t)\right|$ and $E(t)$ obtained with two different bond dimensions ${\chi_1 = 500}$ and ${\chi_2 = 1000}$ for the $12$\hyp site system, and for ${\chi = 250}$ and ${\chi = 500}$ for the $30$\hyp site system, respectively, and also the difference of the respective results. 
For the larger system it was necessary to substantially reduce the values of $\chi$, since otherwise the numerical expenses would exceed the available resources.
We find that the deviation of the results is ${\sim 10^{-6}}$ for the values of $\left|\alpha(t)\right|$ and ${\sim 10^{-4}}$ for the total energy $E(t)$, in the case of the $12$\hyp site system. 
For both observables, this is small compared to the order of magnitude of the observables themselves, so that we conclude these values of $\chi$ suffice to provide quantitatively accurately results, within the dynamical \gls{MPS}+\gls{MF} framework.
For the $30$\hyp site system, however, the deviation is ${\sim 10^{-3}}$ for $\left|\alpha(t)\right|$ and ${\sim 10^{-2}}$ for $E(t)$. 
This is rather large in comparison to the order of magnitude of the observables themselves. 
The data obtained from these calculations is hence only trustworthy in regards to the qualitative physics, but for the larger chain lengths one needs a larger bond dimension to obtain a better quantitative convergence of the results. 
\section{\label{sec:conclusion}Conclusion}
This work presents a self\hyp consistent real\hyp time \gls{MPS}+\gls{MF} approach for investigating the time evolution of a \gls{3D} extended Hubbard model after a fast ramp.  
By combining perturbation theory with a \gls{MF} ansatz, we construct an effective \gls{1D} Hamiltonian \cref{eq:final_eff_MF_Ham} capable of capturing the dynamical build\hyp up of \gls{SC} correlations for this \gls{3D} model system, when quenching or rapidly ramping into a Hamiltonian parameter regime corresponding to \gls{SC} order in equilibrium. 
This approach is generic to any \gls{3D} system composed out of gapped \gls{1D} systems of fermions, as long as coupling between \gls{1D} systems is sufficiently weak for single\hyp fermion tunneling in\hyp between \gls{1D} systems to be suppressed. 
For concrete demonstration of the performance of this approach, we chose systems of \gls{1D} extended Hubbard chains, arranged in parallel in a \gls{2D} square array, forming a \gls{3D} system with weak interchain tunneling $t_\perp$, negative onsite repulsion $U$, and nearest\hyp neighbor interaction $V$ along each chain.
We benchmark the self\hyp consistent algorithm introduced on the simplest possible version \cref{eq:final_eff_MF_Ham_simplified} of the resulting effective \gls{MF} Hamiltonian, only taking onsite pairing into account and neglecting the particle\hyp hole terms \cref{eq:self-cons-param-exchange}. 
We test our approach on systems where each chain is up to ${L=30}$ sites long. 
Using this algorithm we compute the time evolution of the \gls{BCS} order parameter for \gls{SC} order $\alpha(t)$, as a direct indicator of dynamically induced superconductivity. 
The results show that \gls{SC} order sets in after a fast ramp from ${V=0.25}$ to ${V=-0.25}$, where the initial $V$-value realizes an insulating \gls{CDW} state, and the final value would correspond to \gls{SC} order at equilibrium. 
These results are broadly comparable to previous \gls{1D} results~\cite{Paeckel_superconductivity} and represent a best\hyp case scenario, in which double occupancies already present in the \gls{CDW} help to form the non\hyp equilibrium \gls{SC} state after the ramp. 
Performing infinite\hyp size extrapolations and studying the effect of the microscopic initial kernel of \gls{SC} order $\alpha_\mathrm{ini}$ shows that dynamically induced superconductivity is not merely a trivial size effect, but actually present in the thermodynamic limit, and even the smallest yet finite magnitude for $\alpha_\mathrm{ini}$ will result in establishing order within a finite window of time. 
At the same time, we find that resource requirements increase substantially with chain length $L$, but several tens of sites and time frames between one and two orders of magnitude in units of inverse fermion tunneling $t^{-1}$ are accessible already with the modest resources employed for the present proof\hyp of\hyp principle work.
The present work presents multiple avenues for interesting and potentially valuable follow\hyp up work. 
One of these would be to move towards a regime that is physically more realistic as far as solid state systems are concerned, in which the pair\hyp binding energies $\Delta E_p$ would be significantly smaller than in the present work. 
This would entail either lowering $U$, or working directly with a \gls{1D} model offering repulsively mediated pairing, such as a doped two\hyp leg Hubbard ladder~\cite{karakonstantakis2011_Ep,Dolfi2015b}. 
This would require retaining more particle\hyp particle terms \cref{eq:self-cons-param-pairing} than we have done for the present proof\hyp of\hyp principle, as well as incorporating the particle\hyp hole terms \cref{eq:self-cons-param-exchange} into the self\hyp consistent time\hyp evolution step, see \cref{fig:self_cons_tevol}. 
This would be straightforward, as a generic ansatz for the first iteration of these terms is practically imposed by the physics of these \gls{1D} systems. 
As detailed in, e.g., ~\cite{Giamarchi2004}, both class of terms decays with an exponential envelope function characterized by the spin\hyp correlation length, which in turn is easy to obtain from static correlators via \gls{DMRG} calculations for the isolated systems.
With this extension, the present work could stimulate a more direct and fruitful collaboration between theory and experiment on dynamically induced \gls{SC} order in solid state systems. 
Such work would start from either identifying existing materials comprised of many \gls{1D} systems of paired electrons in parallel, with coupling weaker than that pairing, or synthesizing such materials. 
The theory presented in the present work would then allow to closely model any experiments on driving dynamically induced superconductivity in these systems, and thus be much better positioned to ascertain whether some experimental measurement truly is a hallmark of a transient superconducting state, and in turn to propose measurements that would prove the existence of such a state. 
Regarding such a modeling of realistic solid state systems, we point out that \gls{MPS}-based techniques are capable of modeling the equilibrium and dynamical out\hyp of\hyp equilibrium evolution of much more complex \gls{1D} systems than the one studied in the present work. 
This includes coupling to phonon baths~\cite{Brockt2015,Jansen2021,arxiv.2207.08243} and multi\hyp orbital systems~\cite{Kaushal2017}, and for spin systems \gls{MPS}+\gls{MF} techniques have already been used to model experiments of \gls{3D} systems comprised of weakly coupled spin ladders~\cite{Klanjsek2008,Bouillot2011}.
At the same time, we point out that existing experiments on ultracold atomic gases confined in optical lattices offer an invaluable platform to validate the \gls{MPS}+\gls{MF} theory for dynamically induced \gls{SC} states, in both the high\hyp $U$ and the low\hyp $U$ regime. 
Systems with all the essential elements of the set\hyp up of this work --- anisotropic \gls{3D} cubic lattices with ${t_\perp/t \ll 1}$, ${U < 0}$ --- can readily be realized in the laboratory. 
These set-ups would thus allow for a direct one\hyp to\hyp one comparison of theory and experiment. 
Such work would advance the field of out\hyp of\hyp equilibrium many\hyp body dynamics simultaneously on both fronts, as well as establish ultracold atoms as clean, highly controlled model systems of dynamically induced \gls{SC} order.
\paragraph{Acknowledgements}
We acknowledge helpful discussions with Hugo Strand, Sebastian Paeckel, Oscar Gr\aa n\"as.
We acknowledges financial support by the ERC Starting Grant from the European Union's Horizon 2020 research and innovation program under grant agreement No. 758935.
SRM and SM acknowledge funding by the Deutsche Forschungsgemeinschaft (DFG, German Research Foundation) - 217133147/SFB 1073, project B03 and TU Clausthal.
We also acknowledge access to computational resources provided by the GWDG.  
This work also used the Cirrus UK National Tier-2 HPC Service at EPCC (http://www.cirrus.ac.uk) funded by the University of Edinburgh and EPSRC (EP/P020267/1).

\begin{appendix}
\section{\label{sec:appendix_self-cons-GS}Self Consistent Ground State Search}
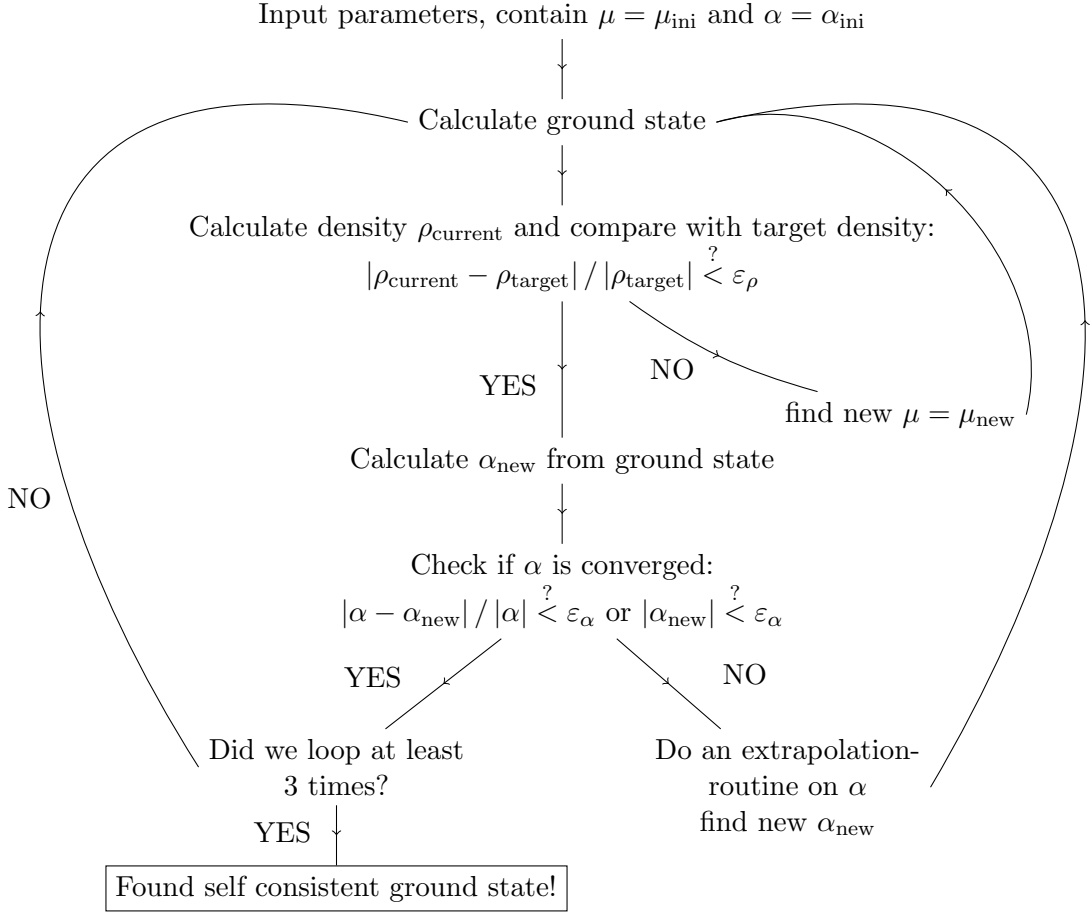
\begin{figure}[!t]
	\centering
	\ifthenelse{\boolean{buildtikzpics}}%
	{%
		%\tikzset{external/export next=false}
\tikzsetnextfilename{self-cons-GS-search_scheme}
\begin{tikzpicture}[
box/.style={rectangle, draw=black, align=center},
ghost/.style={align=center}
]
		\clip (-7.45,0.5) rectangle (7.1,-12);
    \begin{scope}[node distance=0.8cm]
        \node[ghost](input){Input parameters, contain $\mu=\mu_{\text{ini}}$ and $\alpha=\alpha_\text{ini}$};
        \node[ghost](step0)[below=of input]{Calculate ground state};
        \node[ghost](step1)[below=of step0]{Calculate density $\rho_{\text{current}}$ and compare with target density:\\
        $\left|\rho_{\text{current}}-\rho_{\text{target}}\right|/\left|\rho_{\text{target}}\right|\stackrel{?}{<}\varepsilon_{\rho}$};
        \node[ghost](stepNo0)[below=of step1, xshift=4.5cm, yshift = -0.4cm]{find new $\mu=\mu_\text{new}$};
        \node[ghost](step3)[below=of stepNo0, xshift = -4.5cm, yshift = 0.8cm]{Calculate $\alpha_\text{new}$ from ground state};
        \node[ghost](step4)[below=of step3]{Check if $\alpha$ is converged:\\
        $\left|\alpha-\alpha_\text{new}\right|/\left|\alpha\right|\stackrel{?}{<}\varepsilon_\alpha$ or $\left|\alpha_\text{new}\right|\stackrel{?}{<}\varepsilon_\alpha$};
        \node[ghost](stepYes2-0)[below=of step4, xshift=-3cm, yshift=-0.4cm]{Did we loop at least\\
        3 times?};
        \node[box](stepYes3-0)[below=of stepYes2-0, yshift=-0.0cm]{Found self consistent ground state!};
        \node[ghost](stepNo2-0)[below=of step4, xshift=3.0cm, yshift=-0.4cm]{Do an extrapolation-\\
        routine on $\alpha$\\
        find new $\alpha_\text{new}$};
        
        \draw[->-](input) to (step0);
        \draw[->-](step0) to (step1);

        \draw[->-, bend right=10](step1) to (stepNo0) node [left, xshift=-1.5cm, yshift=0.3cm]{NO};
        \draw[->-, bend right = 60](stepNo0.east) to (step0.east);
        \draw[->-](step1) to (step3) node [left, xshift=-0.2cm, yshift=1.0cm]{YES};
        \draw[->-](step3) to (step4);
        \draw[->-](step4) to (stepNo2-0) node [right, xshift=-1.0cm, yshift=1.5cm]{NO};
        \draw[->-](step4) to (stepYes2-0) node [left, xshift=1.0cm, yshift=1.2cm]{YES};
        \draw[->-](stepYes2-0) to (stepYes3-0) node [left, xshift=-0.2cm, yshift=0.75cm]{YES};
        \draw[->-](stepYes2-0.west) .. controls (-8,-5) and (-8,0) .. (step0.west) node[left, xshift=-4.6cm, yshift=-5cm]{NO};
        \draw[->-](stepNo2-0.east) .. controls (8,-5) and (8,0) .. (step0.east);
    \end{scope}
\end{tikzpicture}
	}%
	{%
		\includegraphics{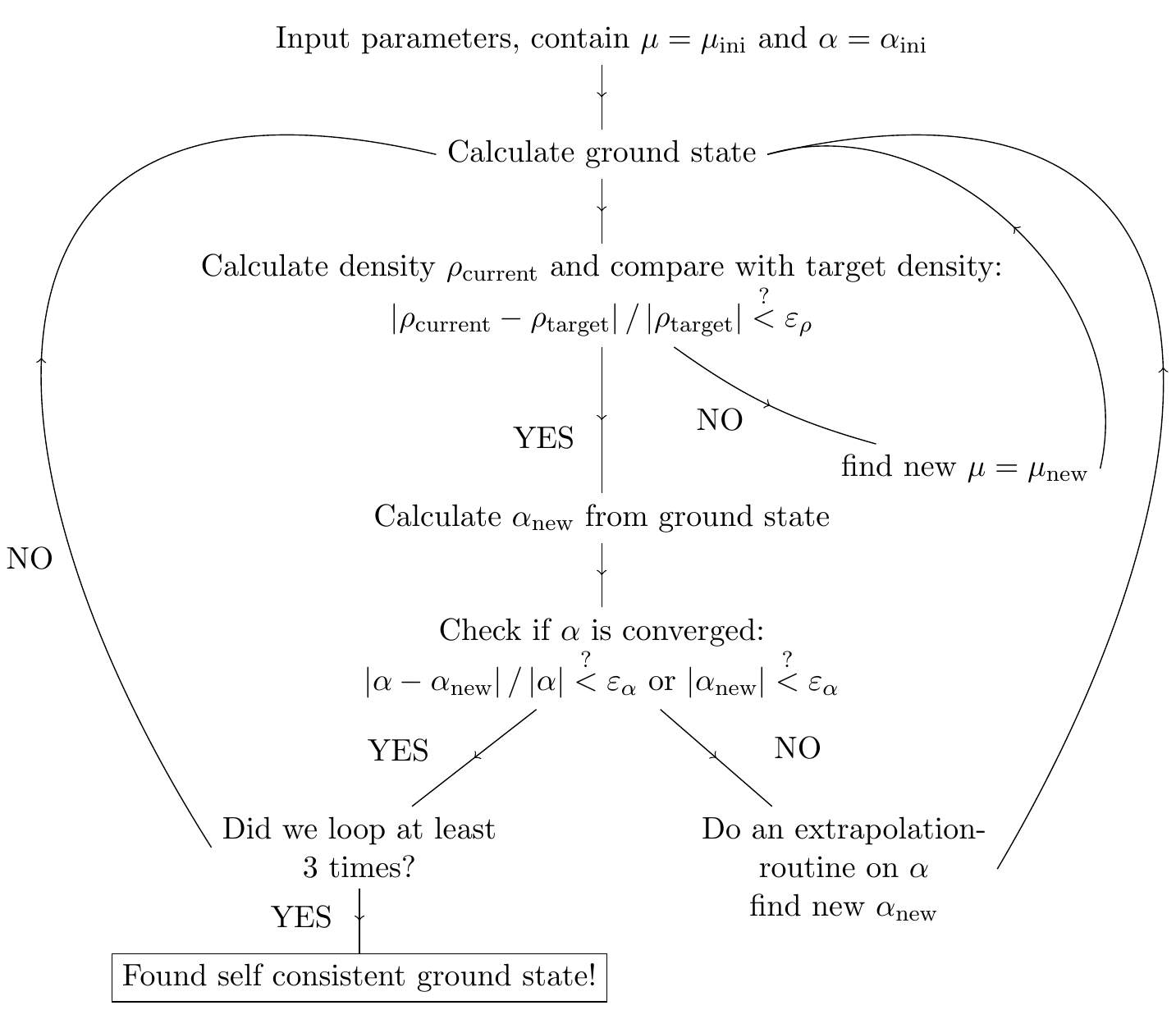}%
	}%
	\caption
	{
		\label{fig:self_cons_gs_search}
		Self consistency loop for the ground-state search. 
		As the MF-parameter $\alpha$ depends on the ground state itself, it has to be adjusted after each DMRG step. 
		As the effective Hamiltonian, furthermore, is no longer particle number conserving we also need to update the chemical potential $\mu$ permanently.
	}
\end{figure}
As mentioned in \cref{sec:resultsAndDiscussion} of this paper, we are making use of the self\hyp consistent ground\hyp state search developed by Bollmark et al.~\cite{Gunnar_coupled_chains, Gunnar_coupled_chains_electrons}. 
Here a brief description of this algorithm shall be given.
Basically, a ground\hyp state search in \gls{MPS} language is an optimization problem solved via \gls{DMRG}. 
However, in our case we are dealing with the special case that not only the state $\ket{\psi}$ has to be optimized but that we also do not know all parameters of the Hamiltonian as one of the parameters, namely $\alpha$, depends on the ground state itself. 
This is why we need to adjust this parameter iteratively during the ground\hyp state search until self consistency is reached, as in any other \gls{MF}\hyp based approach.
By the way $\alpha$ is introduced, the \gls{MF} approximation of our model loses the particle number conservation of the original \gls{3D} Hamiltonian. 
Thus, not only $\alpha$ but also the chemical potential $\mu$ has to be adjusted during the ground\hyp state search. 
At the inception of the iterative procedure $\alpha$ and $\mu$ must be guessed, however crudely. 
Then, we perform a \gls{DMRG}\hyp based ground\hyp state search for this set of parameters, yielding a candidate for a ground state. 
Now, we need to check if the density is at the desired value and if $\alpha$ is consistent. 
First, we measure the density $\rho_\mathrm{current}$ of the state we just calculated and compare it with the density $\rho_\mathrm{target}$ we are targeting. 
If the condition
\begin{align}
    \frac{|\rho_\mathrm{current}-\rho_\mathrm{target}|}{|\rho_\mathrm{target}|} < \varepsilon_\rho\quad \mathrm{with}\quad\varepsilon_\rho\ll1
    \label{eq:condition_convergence_rho_GSsearch}
\end{align}
is fulfilled, we keep the chemical potential $\mu$ we plugged in, if not, a routine that involves interpolation and extrapolation is used to determine a new chemical potential which is applied from this point on.
Second, we measure the value of the \gls{MF} parameter $\alpha$ from the candidate state and check if it is converged via the condition 
\begin{align}
    \left|\alpha_\mathrm{ini}-\alpha_\mathrm{new}\right|/\left|\alpha_\mathrm{ini}\right|<\varepsilon_\alpha\quad \mathrm{or}\quad \left|\alpha_\mathrm{new}\right|<\varepsilon_\alpha \;.
    \label{eq:condition_convergence_alpha_GSsearch}
\end{align}
If this condition is fulfilled, we keep $\alpha$, if not, we once again use a routine that involves extrapolation in order to find a new and better value for $\alpha$. 
Finally, we are either done if both conditions \cref{eq:condition_convergence_rho_GSsearch,eq:condition_convergence_alpha_GSsearch} are fulfilled or we repeat the whole routine using now the new values we obtained for $\alpha$ and $\mu$ as a starting point.
A schematic of the self\hyp consistent ground\hyp state search is depicted in \cref{fig:self_cons_gs_search}.
\section{\label{sec:appendix_rampTime}Effect of the Time Window for the Ramp}
\begin{figure}[!t]
    \centering
	\ifthenelse{\boolean{buildtikzpics}}%
	{%
%		\immediate\write18{paste data/tRamp_0.0/tend_15/tperp_0.2/L_12/U_-4.0/V_-0.25/chi_500/dt_0.01/alphaIni_0.0001/Data_tevol.txt data/tRamp_0.0/tend_15/tperp_0.2/L_12/U_-4.0/V_-0.25/chi_1000/dt_0.01/alphaIni_0.0001/Data_tevol.txt > data/tRamp_0.0/tend_15/tperp_0.2/L_12/U_-4.0/V_-0.25/combined_Data_tevol.txt}
%		\immediate\write18{paste data/tRamp_3.0/tend_15/tperp_0.2/L_12/U_-4.0/V_-0.25/chi_500/dt_0.01/alphaIni_0.0001/Data_tevol.txt data/tRamp_3.0/tend_15/tperp_0.2/L_12/U_-4.0/V_-0.25/chi_1000/dt_0.01/alphaIni_0.0001/Data_tevol.txt > data/tRamp_3.0/tend_15/tperp_0.2/L_12/U_-4.0/V_-0.25/combined_Data_tevol.txt}
%		\tikzset{external/export next=false}
		\tikzsetnextfilename{diffAlpha_difftRamp_tend15_L12_dt0p01_alphaIni0p0001}
		\begin{tikzpicture}
			\begin{groupplot}%
			[%
				group style =%
				{%
					group size			=	2 by 3,%
					vertical sep		=	0.25em,%
					horizontal sep		=	4.5em,%
					x descriptions at	=	edge bottom,%
				},%
				height		=	0.2\textheight,%
				width		=	0.5\textwidth-5.337pt,%
				xlabel		=	{time $t$},%
				xmin		=	0,%
				xmax		=	15,%
				enlarge x limits,
			]%
				\nextgroupplot%
				[%
					ylabel				=	{$\lvert \alpha \rvert / 10^{-5} \vphantom{\rho^{-5}}$},%
					legend pos			=	north west,%
					legend cell align	=	{left},%
					legend style =%
					{%
						draw	=	none,%
						fill	=	none,%
					},%
					scaled y ticks 		=	false,%
					ylabel style 		=	{yshift = -0.2em},%
				]%
					\coordinate (l1) at (rel axis cs:-0.25,0.85);%
					\addplot%
					[%
						color	=	colorA,%
						thick,%
						restrict x to domain	=	0.01:15,%
					]%
						table%
						[%
							x expr = \thisrowno{0},%
							y expr = sqrt(\thisrowno{1}*\thisrowno{1}+\thisrowno{2}*\thisrowno{2})*10^5%
						]%
							{data/tRamp_0.0/tend_15/tperp_0.2/L_12/U_-4.0/V_-0.25/chi_500/dt_0.01/alphaIni_0.0001/Data_tevol.txt};%
					\addlegendentry{$\chi=500$};%
					\addplot%
					[%
						color	=	colorB,%
						thick,%
						restrict x to domain	=	0.01:15,%
					]%
						table%
						[%
							x expr = \thisrowno{0},%
							y expr = sqrt(\thisrowno{1}*\thisrowno{1}+\thisrowno{2}*\thisrowno{2})*10^5%
						]%
							{data/tRamp_0.0/tend_15/tperp_0.2/L_12/U_-4.0/V_-0.25/chi_1000/dt_0.01/alphaIni_0.0001/Data_tevol.txt};%
					\addlegendentry{$\chi=1000$};%				
				\nextgroupplot%
				[%
					ylabel				=	{$\lvert \alpha \rvert  / 10^{-4} \vphantom{\rho^{-5}}$},%
					legend pos			=	north west,%
					legend cell align	=	{left},%
					legend style =%
					{%
						draw	=	none,%
						fill	=	none,%
					},%
					scaled y ticks 		=	false,%
					ylabel style 		=	{yshift = -0.2em},%
				]%
					\coordinate (l2) at (rel axis cs:-0.25,0.85);%
					\addplot%
					[%
						color	=	colorA,%
						thick,%
						restrict x to domain	=	0.01:15,%
					]%
						table%
						[%
							x expr = \thisrowno{0},%
							y expr = sqrt(\thisrowno{1}*\thisrowno{1}+\thisrowno{2}*\thisrowno{2})*10^4%
						]%
							{data/tRamp_3.0/tend_15/tperp_0.2/L_12/U_-4.0/V_-0.25/chi_500/dt_0.01/alphaIni_0.0001/Data_tevol.txt};%
					\addplot%
					[%
						color	=	colorB,%
						thick,%
						restrict x to domain	=	0.01:15,%
					]%
						table%
						[%
							x expr = \thisrowno{0},%
							y expr = sqrt(\thisrowno{1}*\thisrowno{1}+\thisrowno{2}*\thisrowno{2})*10^4%
						]%
							{data/tRamp_3.0/tend_15/tperp_0.2/L_12/U_-4.0/V_-0.25/chi_1000/dt_0.01/alphaIni_0.0001/Data_tevol.txt};%
				\nextgroupplot%
				[%
					ylabel				=	{$\delta_{\lvert \braket{\alpha} \rvert, 500, 1000}$},%
					scaled y ticks 		=	false,%
					ylabel style 		=	{yshift = -0.2em},%
					ymode 				=	log,%
					ymin				=	5e-17,%
					ymax				=	5e-6,%
				]%
					\coordinate (l3) at (rel axis cs:-0.25,0.85);%
					\addplot%
					[%
						color	=	colorC,%
						thick,%
						restrict x to domain	=	0.01:15,%
					]%
						table%
						[%
							x expr = \thisrowno{0},%
							y expr = \eval{abs(sqrt(\thisrowno{1}*\thisrowno{1}+\thisrowno{2}*\thisrowno{2}) - sqrt(\thisrowno{18}*\thisrowno{18}+\thisrowno{19}*\thisrowno{19}))}%
						]%
							{data/tRamp_0.0/tend_15/tperp_0.2/L_12/U_-4.0/V_-0.25/combined_Data_tevol.txt};%
				\nextgroupplot%
				[%
					ylabel				=	{$\delta_{\lvert \braket{\alpha} \rvert, 500, 1000}$},%
					scaled y ticks 		=	false,%
					ylabel style 		=	{yshift = -0.2em},%
					ymode 				=	log,%
					ymin				=	5e-17,%
					ymax				=	5e-6,%
				]%
					\coordinate (l4) at (rel axis cs:-0.25,0.85);%
					\addplot%
					[%
						color	=	colorC,%
						thick,%
						restrict x to domain	=	0.01:15,%
					]%
						table%
						[%
							x expr = \thisrowno{0},%
							y expr = \eval{abs(sqrt(\thisrowno{1}*\thisrowno{1}+\thisrowno{2}*\thisrowno{2}) - sqrt(\thisrowno{18}*\thisrowno{18}+\thisrowno{19}*\thisrowno{19}))}%
						]%
							{data/tRamp_3.0/tend_15/tperp_0.2/L_12/U_-4.0/V_-0.25/combined_Data_tevol.txt};%
				\nextgroupplot%
				[%
					ylabel			=	{$V\vphantom{\rho^{-5}}$},%
					height			=	0.1125\textheight,%
					ylabel style	=	{yshift = -0.2em},%
				]%
					\coordinate (l5) at (rel axis cs:-0.25,0.55);%
					\addplot%
					[%
						color	=	black,%
						thick,%
					]%
						table%
						[%
							x expr = \thisrowno{0},%
							y expr = \thisrowno{1}%
						]%
							{data/tRamp_0.0/tend_15/tperp_0.2/L_12/U_-4.0/V_-0.25/chi_1000/dt_0.01/alphaIni_0.0001/V_in_time.txt};%
				\nextgroupplot%
				[%
					ylabel			=	{$V\vphantom{\rho^{-5}}$},%
					height			=	0.1125\textheight,%
					ylabel style	=	{yshift = -0.2em},%
				]%
					\coordinate (l6) at (rel axis cs:-0.25,0.55);%
					\addplot%
					[%
						color	=	black,%
						thick,%
					]%
						table%
						[%
							x expr = \thisrowno{0},%
							y expr = \thisrowno{1}%
						]%
							{data/tRamp_3.0/tend_15/tperp_0.2/L_12/U_-4.0/V_-0.25/chi_1000/dt_0.01/alphaIni_0.0001/V_in_time.txt};%
			\end{groupplot}%
			\node at (l1) {\subfloat[\label{fig:AbsAlpha_tevol_ramp0}]{}};%
			\node at (l3) {\subfloat[\label{fig:DiffAlpha_tevol_ramp0}]{}};%
			\node at (l5) {\subfloat[\label{fig:V_tevol_ramp0}]{}};%
			\node at (l2) {\subfloat[\label{fig:AbsAlpha_tevol_ramp3}]{}};%
			\node at (l4) {\subfloat[\label{fig:DiffAlpha_tevol_ramp3}]{}};%
			\node at (l6) {\subfloat[\label{fig:VAlpha_tevol_ramp3}]{}};%
		\end{tikzpicture}
	}%
	{%
		\includegraphics{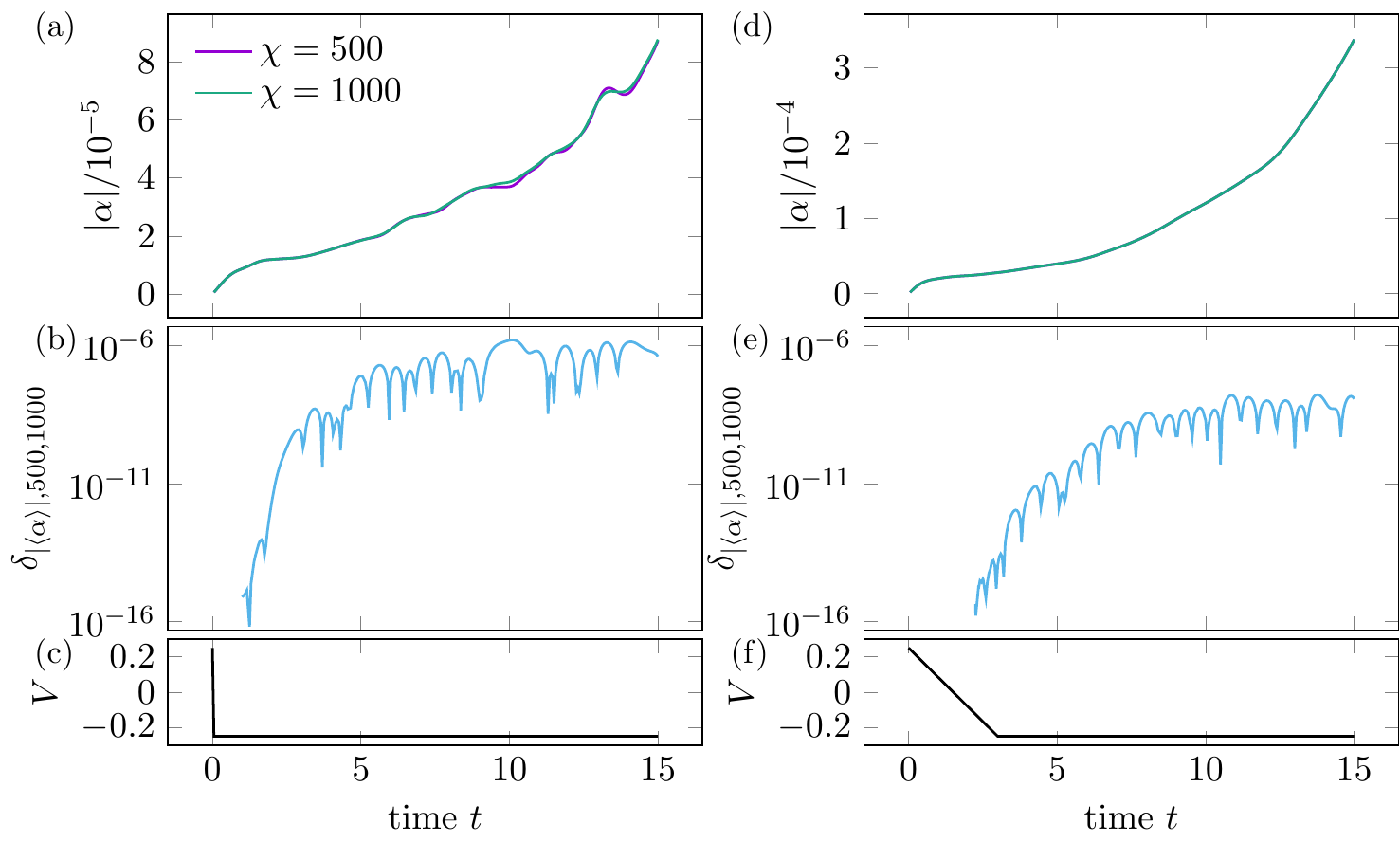}

		{\phantomsubcaption\label{fig:AbsAlpha_tevol_ramp0}}%
		{\phantomsubcaption\label{fig:DiffAlpha_tevol_ramp0}}%
		{\phantomsubcaption\label{fig:V_tevol_ramp0}}%
		{\phantomsubcaption\label{fig:AbsAlpha_tevol_ramp3}}%
		{\phantomsubcaption\label{fig:DiffAlpha_tevol_ramp3}}%
		{\phantomsubcaption\label{fig:VAlpha_tevol_ramp3}}%
	}%
    \caption
    {
	    \label{fig:Diffalpha_tevol_ramp3.0vs0.0}
		Difference between the data of the time evolution of $\left|\alpha\right|$ calculated for two different bond dimensions ${\chi = 500}$ and ${\chi = 1000}$ in a $12$\hyp site system for two different ramp times $\Delta t_\mathrm{ramp}$. 
    	We gain an accuracy of the order of $10^2$ via increasing the ramp time from 0.0 to 3.0. 
	}
\end{figure}
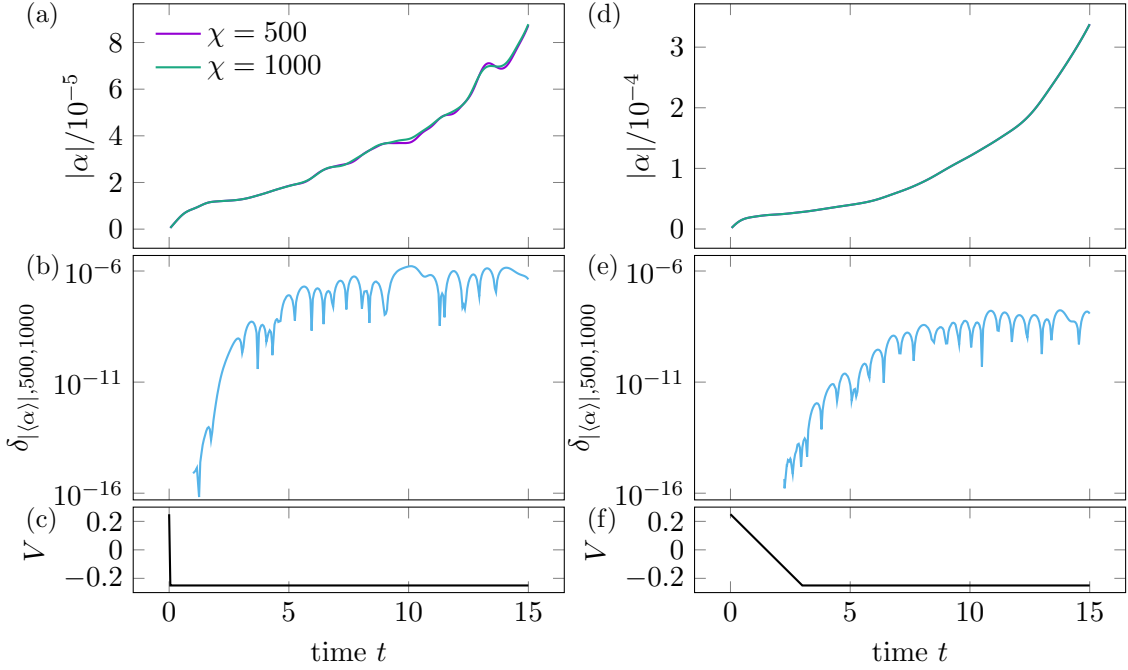
In \cref{sec:resultsAndDiscussion} it was mentioned that a ramp appeared to be numerically more stable than an instantaneous quench. 
For a more detailed explanation of this statement, we compare the accuracy of the data we measure for the \gls{MF} parameter $\left|\alpha\right|$ for a quench and a ramp in \cref{fig:Diffalpha_tevol_ramp3.0vs0.0}.
Changing $V$ either through an instantaneous quench or through a fast continuous ramp, which we have used throughout the main text, we evolve our system up to times of ${t_\mathrm{end} = 15}$. 
In both cases we compare the variance between the $\alpha$ data for two different bond dimensions $\chi$, as it was done in \cref{sec:accuracyOfData} as a check of accuracy. 
We find that difference is two orders of magnitude smaller for the ramp compared to the case of the instantaneous quench. 
This is why we chose to use ramps for all our calculations presented in this paper.
\end{appendix}
\bibliography{Literatur}
\end{document}